\documentclass[aps,prd]{revtex4}
\usepackage{epsfig,epsf}
\usepackage{amsmath}
\usepackage{amsthm}
\usepackage{amsfonts}
\usepackage{amssymb}
\usepackage{dsfont}
\usepackage{multirow}
\usepackage{appendix}
\usepackage{slashed}
\usepackage[active]{srcltx}
\usepackage{psfrag}
\usepackage{subfigure}
\usepackage[colorlinks,citecolor=blue,urlcolor=red,linkcolor=red]{hyperref}
\usepackage[colorlinks,citecolor=blue,urlcolor=red,linkcolor=red]{hyperref}
\begin{document}

\title{{\Large{\bf  $D D A$,  $D^{*}D^{*}A$ and $D^{*} D A$ vertices in the light-cone QCD and  considering  $B^0\to {K}_{1}^{+} \pi^-$ branching ratio }}}

\author{\small
S. Momeni \footnote {e-mail: samira.momeni@ph.iut.ac.ir }, R.
Khosravi   \footnote {e-mail: rezakhosravi@iut.ac.ir }}

\affiliation{\emph{Department of Physics, Isfahan University of
Technology, Isfahan 84156-83111, Iran} }

\begin{abstract}
We investigate the strong coupling constants of $D D A$,
$D^{*}D^{*}A$ and $D^{*} D A$ vertices in the framework of the
light-cone QCD sum rules, where $A$ is an axial vector meson such as
$a_1, b_1, K_{1A}, K_{1B}, K_1(1270)$ and $K_1(1400)$. Using the
strong coupling constants of  $D_s D K_1$, $D_s^* D K_1$ and $D_s^*
D^* K_1$ vertices for $K_1(1270)$ and $K_1(1400)$ mesons, we
evaluated the branching ratios of the non-leptonic decays $B^0\to
{K}_{1}^{+}(1270,1400) \pi^-$. Our results for the branching ratios
of these decays are a good agreement with the experimental values.
\end{abstract}


\maketitle

\section{Introduction}
The strong coupling constants are very useful resources for
understanding the nature of the strong interactions and hadronic
phenomena. The strong couplings of the charmed mesons have a
significant role in the hadronic decays of $B$ meson when
phenomenological models, for instance one-particle exchange model,
are used. The phenomenological Lagrangian of  the one-particle
exchange model for the hadronic decays of $B$ meson contains input
parameters such as $\beta$ and $\lambda$, which describe the strong
couplings  connected to  the charmed mesons in these decays
\cite{HQEFT}. Therefore, the calculation of the strong form factors
and coupling constants, especially vertices composed of the charmed
mesons, has attracted much attention. Until now, researchers have
computed some coupling constants of the charmed mesons such as $D^*
D^* \rho$ \cite{MEBracco}, $D^* D \pi$ \cite{FSNavarra,MNielsen}, $D
D \rho$ \cite{MChiapparini}, $D^* D \rho$ \cite{Rodrigues3}, $D D
J/\psi$ \cite{RDMatheus},  $D^* D J/\psi$ \cite{RRdaSilva},
$D^*D_sK$, $D^*_sD K$, $D^*_0 D_s K$, $D^*_{s0} D K$ \cite{SLWang},
$D^*D^* P$, $D^*D V$, $D D V$ \cite{ZGWang}, $D^* D^* \pi$
\cite{FCarvalho}, $D_s D^* K$, $D_s^* D K$ \cite{ALozea}, $D D
\omega$ \cite{LBHolanda},  $D_s D_s V$, $D^{*}_s D^{*}_s V$
\cite{KJ,KJ2}, and $D_1D^*\pi, D_1D_0\pi, D_1D_1\pi$ \cite{Janbazi}.
These coupling constants are often evaluated within the framework of
the QCD sum rules.

In this work, we decide to calculate the strong coupling constants
associated with $D D a_1$, $D D b_1$, $D_s D K_{1A}$, $D_s D
K_{1B}$, $D^* D a_1$, $D^* D b_1$, $D_s^* D K_{1A}$, $D_s^* D
K_{1B}$, $D^* D^* a_1$, $D^* D^* b_1$, $D_s^* D^* K_{1A}$ and $D_s^*
D^* K_{1B}$ vertices in the frame work of the light-cone sum rules
(LCSR). Also, the strong coupling constants related to $D_s D K_1$,
$D_s^* D K_1$ and $D_s^* D^* K_1$ vertices for $K_1(1270)$ and
$K_1(1400)$ axial vector mesons are estimated by the corresponding
vertices of $K_{1A}$ and $K_{1B}$ mesons. For example, the relations
for the coupling constants $g_{D_s D K_1(1270)}$ and $g_{D_s D
K_1(1400)}$ are as
\begin{eqnarray}\label{eq.01}
g_{D_s D K_1(1270)}&=&g_{D_s D K_{1A}}\,\sin\theta_K
+g_{D_s D K_{1B}}\,\cos\theta_K\,,\nonumber\\
g_{D_s D K_{1}(1400)}&=&g_{D_s D K_{1A}}\,\cos\theta_K -g_{D_s D
K_{1B}}\,\sin\theta_K\,,
\end{eqnarray}
where $\theta_K$ is the mixing angle. Similar expressions can be
written for $f_{D_s^* D K_{1}(1270,1400)}$ and $h_{D_s^* D^*
K_{1}(1270,1400)}$.

As an example of specific application of these coupling constants
can be pointed out to branching ratio calculations of hadronic $B$
decays. In this paper, we would like to consider the branching
ratios of the decays $$B^0\to {K}_{1}^{+}(1270, 1400) \pi^-\,,$$
according to the coupling constants of $D_s D K_1$, $D_s^* D K_1$
and $D_s^* D^* K_1$ vertices.

The plan of the present paper is as follows: In section II, the
strong coupling constants $g_{D D A}$, $f_{D^* D A}$ and $h_{D^* D^*
A}$ are calculated in the framework of the LCSR.   In section III,
we analyze and estimate the strong coupling constants for the
aforementioned vertices.  In addition, we consider the branching
ratio of $B^0\to {K}_{1}^{+} \pi^-$ decay for $K_1(1270)$ and
$K_1(1400)$ mesons using the coupling constants of $D_s D K_1$,
$D_s^* D K_1$ and $D_s^* D^* K_1$ vertices and compare our results
with the experimental values and predictions of other methods.

\section{Strong coupling constants in the LCSR}
In the LCSR, the strong coupling constants $g_{D_{(s)} D A}$,
$f_{D^*_{(s)} D A}$ and $h_{D^*_{(s)} D^* A}$ are evaluated with
$\Pi^{D_{(s)} D A}$, $\Pi^{D^*_{(s)} D A}_{\mu}$ and $\Pi^{D^*_{(s)}
D^* A}_{\mu\nu}$ correlation functions, respectively. From now on
for simplicity, we use $D (D^*)$ instead of  $D^0 (D^{*0}), D^+
(D^{*+})$ and $D_s (D^*_s)$ in our formulations. The aforementioned
correlation functions are defined as
\begin{eqnarray}\label{eq.1}
\Pi^{D D A }(p,q)&=&i \int d^4x \, e^{-i q \cdot x} \langle 0
|\mathcal{T}\{j^{D}(0) \,{j^{D}}^{\dagger}(x)\}|A(p)\rangle, \nonumber\\
\Pi^{D^* D A}_{\mu}(p,q)&=&i \int d^4x \, e^{-i q \cdot x} \langle 0
|\mathcal{T}\{j^{D^*}_\mu(0)\, {j^D}^{\dagger}(x)\}|A(p)\rangle,
\nonumber\\
\Pi^{D^* D^* A}_{\mu\nu}(p,q)&=&i \int d^4x \, e^{-i q \cdot x}
\langle 0 |\mathcal{T}\{j^{D^*}_\mu(0)\,
{j_\nu^{D^*}}^{\dagger}(x)\}|A(p)\rangle,
\end{eqnarray}
where  $\mathcal{T}$ is the time-ordering operator. In addition,
$j^{D}=i \bar{q_{i}}(1-\gamma_5)c$ and $j^{D^*}_\mu=i\bar{q_{i}}
\gamma_{\mu} c$  ($q_{i}$ is the field of the light quark from which
the charmed meson is made; $u, d$, or $s$) are the interpolating
currents for $D$ and $D^*$ mesons, respectively. The main reason for
choosing the Chiral current $i \bar{q_{i}}(1-\gamma_5)c$ for $D$
meson instead of the usual pseudoscalar current $i
\bar{q_{i}}\gamma_5 c$ is to provide the results with less
uncertainties \cite{Huang}.

In the LCSR approach, the correlation functions $\Pi^{D D A}$,
$\Pi^{D^* D A}_{\mu}$ and $\Pi^{D^* D^* A}_{\mu\nu}$ can be
calculated in two different ways. In  the physical or
phenomenological representation and the QCD or theoretical ones. The
strong coupling constants $g_{D D A}$, $f_{D^* D A}$ and $h_{D^* D^*
A}$ can be obtained by using the dispersion relation to link these
two representations of the correlation functions.

\subsection{The phenomenological side}
In the phenomenological part, $D D A$, $D^* D A$ and $D^* D^* A$
vertices can be studied in terms of hadronic parameters. To obtain
the phenomenological side  of the correlation functions, we can
insert two complete sets of intermediate states with the same
quantum numbers as the meson currents into these correlation
functions. After isolating the higher-state contributions from the
pole terms of charmed mesons and performing the Fourier
transformation, we have:
\begin{eqnarray}\label{eq028}
\Pi^{D D A}(p,q)&=&\frac{\langle 0 |j^D|D(p+q)\rangle\,\langle D
(p+q) |A(p)\,D(q)\rangle\, \langle D(q) |{j^{D}}^{\dagger}|0
\rangle}{(m^2_D-q^2)\,[m^2_D-{(p+q)}^2]} +\mbox{higher
and continuum states}\,,\nonumber\\
\Pi^{D^* D A}_{\mu}(p,q)&=&\frac{\langle 0
|j_{\mu}^{D^{*}}|D^{*}(p+q)\rangle\,\langle D^{*} (p+q)
|A(p)\,D(q)\rangle\, \langle D(q) |{j^{D}}^{\dagger}|0
\rangle}{(m^2_D-q^2)\,[m^2_{D^{*}}-{(p+q)}^2]}+\mbox{higher
and continuum states}\,,\nonumber\\
\Pi^{D^* D^* A}_{\mu\nu}(p,q)&=&\frac{\langle 0
|j_{\mu}^{D^*}|D^*(p+q)\rangle\,\langle D^* (p+q)
|A(p)\,D^{*}(q)\rangle\, \langle D^{*}(q) |{j_\nu^{D^*}}^{\dagger}|0
\rangle}{(m^2_{D^{*}}-q^2)\,[m^2_{D^*}-{(p+q)}^2]}+\mbox{higher and
continuum states}\,.
\end{eqnarray}
Using the following matrix elements:
\begin{eqnarray}\label{eq.3}
\langle 0 |j^D(0)|D(p+q)\rangle&=&\frac{f_D\,m_{D}^2}{m_c+m_{q_i}}\,,\nonumber\\
\langle D (p+q) |A(p)\,D(q)\rangle&=&2\,g_{D D A}\,\varepsilon.q\,,\nonumber\\
\langle 0 |j^{D^*}_\mu(0)|D^*(p+q)\rangle&=& f_{D^*}\,m_{D^*}\,\varepsilon^*_\mu\,,\nonumber\\
\langle D^{*}(p+q) |A(p)\,D(q)\rangle&=&4i\, f_{D^* D
A}\,\epsilon^{\alpha\beta\sigma\lambda}\,p_{\alpha}q_{\beta}
\varepsilon^*_{\sigma}\varepsilon_{\lambda}\,,\nonumber\\
\langle D^*(p+q) |A(p)\,D^*(q)\rangle&=&i~h_{D^*D^* A} \left.[
(p+2q)^{\alpha} g^{\beta\lambda}+ (p+q)^{\beta} g^{\lambda\alpha} +
q^{\lambda} g^{\alpha\beta}\right.]\,\varepsilon^*_{\alpha}\,
\varepsilon_{\beta}\, \varepsilon^*_{\lambda}(q)\,,
\end{eqnarray}
where $\varepsilon^*$, $\varepsilon$ and $\varepsilon^*(q)$
represent the polarizations of $D^*(p+q)$, $A$ and $D^*(q)$ mesons
respectively, the following results are obtained:
\begin{eqnarray}\label{eq.4}
\Pi^{D D
A}(p,q)&=&\frac{2\,f_D^2\,m_D^4}{{(m_c+m_{q_{i}})}^2(m^2_D-q^2)\,[m^2_D-{(p+q)}^2]}\,g_{D
D A}\,\varepsilon.q+\mbox{higher and continuum states}\,, \nonumber\\
\Pi^{D^* D A}_{\mu}(p,q)&=&\frac{4
i\,f_{D}\,f_{D^*}\,m_D^2\,m_{D^*}}{(m_{c}+m_{q_{i}})
(m^2_D-q^2)\,[m^{2}_{D^{*}}-{(p+q)}^2]} \,f_{D^* D A}\,
\epsilon_{\mu\lambda\alpha\beta}\,\varepsilon^{\lambda}p^{\alpha}q^{\beta}
+\mbox{higher and continuum states}\,,\nonumber\\
\Pi^{D^* D^* A}_{\mu\nu}(p,q)&=&
\frac{i\,f^2_{D^*}\,m^2_{D^*}}{(m^2_{D^*}-q^2)\,[m^2_{D^*}-{(p+q)}^2]}\,
h_{D^* D^* A}\, (p+q)_{\mu} \varepsilon_{\nu} +\mbox{higher and
continuum states}\,,
\end{eqnarray}
where $g_{D D A}$, $f_{D^* D A}$ and $h_{D^* D^* A}$ are the strong
coupling constants, $m_D$, $m_{D^*}$ and  $f_D$, $f_{D^*}$ are
masses and decay constants of mesons, respectively.  Any arbitrary
structure in the correlation function can be selected to compute the
strong coupling. Here, calculations are done for the Lorentz
structures $\varepsilon.q$,
$\epsilon_{\mu\lambda\alpha\beta}\,\varepsilon^{\lambda}p^{\alpha}q^{\beta}$
and $(p+q)_{\mu} \varepsilon_{\nu}$ from  $\Pi^{D D A}$, $\Pi^{D^* D
A}_{\mu}$  and  $\Pi^{D^* D^* A}_{\mu\nu}$, respectively.

\subsection{The theoretical side}
To calculate the QCD or the theoretical part of $\Pi^{D D A}$,
$\Pi^{D^* D A}_{\mu}$  and  $\Pi^{D^* D^* A}_{\mu\nu}$ in the LCSR
approach, the $\mathcal{T}$ product of the interpolating currents
should be expanded at the light-like distances $x^{2}\simeq 0$.
After contracting the $c$ quark field, the correlation functions
\begin{eqnarray}\label{eq.5}
\Pi^{D D A}(p,q)&=&- \int d^4x \, e^{-i q \cdot x} \, \langle 0
|{\bar {q_{i}}}
(0)\,(1-\gamma_5)\,S_{c}(x,0)\,(1-\gamma_5)\,q_{j}(x)|A(p)\rangle
\, , \nonumber\\
\Pi^{D^* D A}_{\mu}(p,q)&=&- \int d^4x \, e^{-i q \cdot x} \,
\langle 0 |{\bar {q_{i}}}
(0)\,\gamma_{\mu}\,S_{c}(x,0)\,(1-\gamma_5)\,q_{j}(x)|A(p)\rangle \,
,
\nonumber\\
\Pi^{D^* D^* A}_{\mu\nu}(p,q) &=&- \int d^4x \, e^{-i q \cdot x} \,
\langle 0 |{\bar {q_{i}}}
(0)\,\gamma_{\mu}\,S_{c}(x,0)\,\gamma_{\nu}\,q_{j}(x)|A(p)\rangle
\,,
\end{eqnarray}
are obtained. In these phrases, $S_c(x, 0)$ is the propagator of $c$
quark, $q_i$ and $q_j$ are the fields of the light quarks that are
located inside the two charmed mesons. For calculating the
theoretical part of the correlation function, the Fierz
rearrangement is utilized. As a result of the Fierz rearrangement,
the combination of $\Gamma^\lambda\Gamma_\lambda$ is appeared before
$q_{j}(x)$ in the correlation functions, where $\Gamma_\lambda$ is
the full set of the Dirac matrices, $\Gamma_\lambda = (I, \gamma_5,
\gamma_\mu, \gamma_\mu\gamma_5, \sigma_{\mu\nu} )$. After
rearrangement the quantum fields and matrices, the correlation
functions turn into two parts including a trace section and a matrix
element of non–local operators between $A$ meson and vacuum state,
as
\begin{eqnarray}\label{eq033}
\Pi^{D D A}(p,q)&=&\frac{i}{4}\,\int d^4x\int
\frac{d^4k}{(2\pi)^4}\frac{e^{i(k-q).x}}{k^{2}-m_{c}^{2}}
~\mbox{Tr}~ (1-\gamma_{5})(\not k+m_{c} )(1-\gamma_{5})
\,\Gamma^{\lambda}\,\langle 0| \bar{q_i}(0)\,\Gamma_{\lambda}\,
q_j (x) | A (p)\rangle,\nonumber\\
\Pi^{D^* D A}_{\mu}(p,q)&=&\frac{i}{4}\,\int d^4x\int
\frac{d^4k}{(2\pi)^4}\,\frac{e^{i(k-q).x}}{k^{2}-m_{c}^{2}}
~\mbox{Tr}~ \gamma_{\mu}(\not k+m_{c} )(1-\gamma_{5})
\,\Gamma^{\lambda}\langle 0| \bar{q_i}(0)\,\Gamma_{\lambda}\,
q_j (x) | A (p)\rangle,\nonumber\\
\Pi^{D^* D^* A}_{\mu\nu}(p,q)&=&\frac{i}{4}\,\int d^4x\int
\frac{d^4k}{(2\pi)^4}\,\frac{e^{i(k-q).x}}{k^{2}-m_{c}^{2}}
~\mbox{Tr}~ \gamma_{\mu}(\not k+m_{c} )\gamma_{\nu}
\,\Gamma^{\lambda}\langle 0| \bar{q_i}(0)\,\Gamma_{\lambda}\, q_j
(x) | A (p)\rangle.
\end{eqnarray}
In the LCSR approach the non-zero matrix elements $\langle 0|
\bar{q_i}(0)\,\Gamma_{\lambda}\, q_j (x) | A (p)\rangle$, called the
light-cone distribution amplitudes (LCDAs), are defined in terms of
twist functions. For instance, the two-parton distribution amplitude
for the axial vector meson $A$, with the light quark content $q_i$
and $q_j$, is given as \cite{Kwei}:
\begin{eqnarray}\label{eq34}
\langle 0| \bar{q_i}^{\alpha}(0)\,q_j^\delta (x) | A (p,\varepsilon)
\rangle &=& -\frac{i}{4}  \int_0^1 du~ e^{-i u  p. x}\Bigg\{ f_{A}
m_{A} \Bigg[ \not\! p\gamma_5 \frac{\varepsilon. x}{p.x}
\Phi_\parallel(u) +\Bigg( \not\! \varepsilon -\not\! p
\frac{\varepsilon. x}{p.x}\Bigg)\gamma_5 g_\perp^{(a)}(u) \nonumber\\
&-& \not\! x\gamma_5 \frac{\varepsilon. x}{2(p.x)^2} m_{A}^2
\phi_{b}(u) + \epsilon_{\mu\nu\rho\sigma} \varepsilon^{\nu}
p^{\rho} x^\sigma \gamma^\mu
\frac{g_\perp^{(v)}(u)}{4}\Bigg] \nonumber\\
&+& \,f^{\perp}_{A} \Bigg[ \frac{1}{2}( \not\! p\not\!\epsilon-
\not\!\epsilon \not\! p ) \gamma_5\,\Phi_\perp(u) - \frac{1}{2}(
\not\! p\not\! x- \not\! x \not\! p ) \gamma_5 \frac{\varepsilon.
x}{(p.x)^2} m_{A}^2 \bar
h_\parallel^{(t)} (u) \nonumber\\
&+& i \Big(\varepsilon. x\Big) m_{A}^2 \gamma_5
\frac{h^{(p)}_\parallel (u)}{2} \Bigg]\Bigg\}^{\delta\alpha},
\end{eqnarray}
where  $\Phi_\parallel$, $\Phi_\perp$ are twist-2, $g_\perp^{(a)}$,
$g_\perp^{(v)}$, $h_\parallel^{(t)}$ and $h_\parallel^{(p)}$ are
twist-3  functions. In addition, $\phi_{b}$ and $\bar
h_\parallel^{(t)}$ are defined as
\begin{eqnarray*}\label{eq35}
\phi_{b} (u) &=&\Phi_\parallel(u) -2 g_\perp^{(a)}(u),\nonumber\\
\bar h_\parallel^{(t)}(u) &=& h_\parallel^{(t)}(u)- \frac{1}{2}
\Phi_\perp(u),
\end{eqnarray*}
for $x^{2}\neq 0$. Moreover, $f_{A}$ and $f^{\perp}_{A}$ are the
decay constants of the axial vector meson $A$.  The explicit
expressions for the relevant two-parton distribution amplitudes and
definitions for the above mentioned twist functions are collected in
Appendix.

After substituting the two-parton distribution amplitudes of the
axial vector meson $A$ into the correlation functions $\Pi^{D D
A}(p,q)$, $\Pi^{D^* D A }_{\mu}$ and $\Pi^{D^* D^* A}_{\mu\nu}$, we
should calculate some traces and then integrals over  variables $x$
and $u$. In the next step,  we equate the coefficients of the
structures $\varepsilon.q$,
$\epsilon_{\mu\lambda\alpha\beta}\,\varepsilon^{\lambda}p^{\alpha}q^{\beta}$
and $(p+q)_{\mu} \varepsilon_{\nu}$ from both the phenomenological
and theoretical sides of $\Pi^{D D A}(p,q)$, $\Pi^{D^* D A }_{\mu}$
and $\Pi^{D^* D^* A}_{\mu\nu}$, respectively. Finally, to apply the
Borel transformations with respect to two variables ${(p+q)}^2$ and
$q^2$, the strong couplings $g_{D D A}$, $f_{D^* D A}$ and $h_{D^*
D^* A}$ are obtained in the LCSR as
\begin{eqnarray}\label{eq335}
g_{D D A} &=&\frac{\Delta(u_{0},
s_{0})f_{A}^{\perp}}{8\delta_{1}}\exp\left[ \frac{m^2_{D}}{M_1^2}+
\frac{m^2_{D}}{M_2^2}\right]
\left[ m_{A}^{2}m_{c}(2-u_{0})h_\|^{(p)}(u_{0}) \right],\nonumber\\
f_{D^*D A}&=&\frac{\Delta(u_{0},
s_{0})M_0^{2}f_{A}}{8\delta_{2}}\exp\left[
\frac{m^{2}_{D^{*}}}{M_{1}^2}+
\frac{m^{2}_{D}}{M_{2}^2}\right]\left[
5\frac{f_{A}^{\perp}}{f_{A}}\,\Phi_\perp(u_{0})+
\frac{m_{A}m_{c}}{M_0^{2}}g_\perp^{(v)}(u_{0}) \right],\nonumber\\
h_{D^* D^* A}&=&\frac{\Delta(u_{0},
s_{0})M_0^{2}f_{A}^{\perp}\,m_{c}}{\delta_{3}}\exp\left[
\frac{m^{2}_{D^{*}}}{M_{1}^2}+
\frac{m^{2}_{D^{*}}}{M_{2}^2}\right]\left[ 2\Phi_\perp(u_{0})+
\frac{m_{A}^{2}}{M_0^{2}}\bar h_\parallel^{(t)(ii)}(u_{0}) \right],
\end{eqnarray}
where
\begin{eqnarray*}
\begin{array}{ll}
\delta_{1}=\frac{f_{D}^2 m_{D}^4}{(m_c+m_{q_{i}})\,(m_c+m_{q_{j}})},
&
\delta_{2}=\frac{f_{D^{*}} f_{D} m_{D^{*}} m_{D}^2}{(m_c+m_{q_{i}})},\\
\delta_{3}=f^2_{D^{*}}m^2_{D^{*}},     &
u_{0}=\frac{M_{1}^2}{M_{1}^2+M_{2}^2},                                                \\
M_0^{2}=\frac{M_{1}^2\,M_{2}^2}{M_{1}^2+M_{2}^2},                  &
\bar h_\parallel^{(t)(ii)}(u)=\int_0^u dv\int_0^v d\omega~ \bar
h_\parallel^{(t)}(\omega),                                                            \\
\Delta(u_{0},
s_{0})=\exp\left[-\frac{u_{0}\,(1-u_{0})\,m_{A}^2+m_{c}^2}{M_0^2}\right]-\exp\left[
-\frac{s_{0}}{M_0^2}\right].
\end{array}
\end{eqnarray*}
Parameter $s_0$ is the continuum threshold that appears in function
$\Delta(u_{0}, s_{0})$. To calculate the coupling constant of vertex
$D D A (D_s D A)$ in Eq. (\ref{eq335}), the continuum threshold
$s_0$ is connected to $D (D_s)$. On the other hand, in vertices
$D^*D A (D_s^*D A)$ and $D^* D^* A (D_s^* D^* A)$, it is related to
$D^* (D_s^*)$ meson. The continuum thresholds for the charmed mesons
have been calculated via the QCD sum rules (SR) in Ref.
\cite{Hayashigaki2004}. Their results for $\sqrt{s_0}$ are presented
in Table \ref{Ta11}.
\begin{table}[th]
\caption{The continuum threshold parameters for $D$, $D_s$, $D^*$
and $D_s^*$ mesons in \rm{GeV}.} \label{Ta11}
\begin{ruledtabular}
\begin{tabular}{ccccc}
Meson &$D$&$D_s$&$D^*$&$D_s^*$\\
\hline
$\sqrt{s_0}$&$2.45\pm0.15$&$2.50\pm0.20$&$2.55\pm0.05$&$2.56\pm0.15$\\
\end{tabular}
\end{ruledtabular}
\end{table}

The Borel transformations, used in extracting the strong coupling
relations in Eq. (\ref{eq335}), are as follows.
\begin{eqnarray*}
B_{\Lambda^2}(M^2)\left[\frac{1}{\Lambda^2-m^{2}}\right]=-\frac{e^{-\frac{m^{2}}{M^{2}}}}
{M^{2}}\,,~~~~~~~~~~~
B_{\Lambda^2}(M^2)~\left[e^{-\alpha\,\Lambda^2}\right]=\delta~(1-\alpha\,M^2)\,,
\end{eqnarray*}
where $\Lambda$ is variable. In the Borel transformations, parameter
$M$ is known as the Borel mass.

Now,  the values of the strong couplings $g_{D D a_1}$, $g_{D D
b_1}$, $g_{D_s D K_{1A}}$, $g_{D_s D K_{1B}}$, $f_{D^* D a_1}$,
$f_{D^* D b_1}$, $f_{D_s^* D K_{1A}}$, $f_{D_s^* D K_{1B}}$, $h_{D^*
D^*a_1}$, $h_{D^* D^* b_1}$, $h_{D_s^* D^* K_{1A}}$ and $h_{D_s^*
D^* K_{1B}}$ can be estimated numerically, with the help of Eq.
(\ref{eq335}) obtained using the LCSR. In addition, the results of
the strong couplings $g_{D_{s} D K_{1}}$, $f_{D_s^* D K_{1}}$ and
$h_{D_s^* D^* K_{1}}$ for $K_1(1270)$ and $K_1(1400)$ mesons are
calculated by Eq. (\ref{eq.01}).

\section{Numerical analysis}
In this section, our numerical analysis is presented for the strong
coupling constants $g_{{D D A}}$, $f_{{D^{*} D A}}$ and $h_{{D^{*}
D^{*} A}}$. In this work, the masses of the light quarks $u$ and $d$
are neglected. The masses for $s$ and $c$ quarks are taken in
$\mbox{GeV}$ as $m_{s}=(0.09\pm 0.00)$ and $m_c=(1.28\pm 0.03)$,
respectively \cite{pdg}. For charmed mesons, the masses are used in
$\mbox{GeV}$ as $m_{D}=1.86  $,  $m_{D_s}= 1.96 $, $m_{D^*}=2.01$
and $m_{D^{*}_s}= 2.11$ \cite{pdg}. In this work, the decay constant
values of the charmed  mesons $D^{(*)}$ and $D^{(*)}_s$, obtained in
the SR, are used  as $f_{D}=(201 \pm{13})~\mbox{MeV}$, $f_{D_s}=
(238 \pm 23)~\mbox{MeV}$, $f_{D^*}=(242 \pm{20})~\mbox{MeV}$ and
$f_{D^*_s}= (314 \pm 19)~\mbox{MeV}$ \cite{Mutuk}. The decay
constant values of the axial vector mesons, i.e. $f_{A}$ and
$f^\perp_{A}$ are equal at energy scale $\mu=1\,\rm{GeV}$. The
masses and the decay constant values for the axial vector mesons,
evaluated from the LCSR \cite{Kwei}, are presented in Table
\ref{Ta1}.
\begin{table}[th]
\caption{Masses and decay constant values for axial vector mesons
$a_1, b_1, K_{1A}, K_{1B}$.} \label{Ta1}
\begin{ruledtabular}
\begin{tabular}{ccccc}
Axial vector meson &$a_1$&$b_1$&$K_{1A}$&$K_{1B}$\\
\hline
$m_{A}\,\rm(GeV)$&$1.23\pm0.40$&$1.23\pm0.32$&$1.31\pm0.06$&$1.34\pm0.08$\\
$f_{A}\,\rm(MeV)$&$238\pm10$&$180\pm8$&$250\pm13$&$190\pm10$\\
\end{tabular}
\end{ruledtabular}
\end{table}

\subsection{ Analysis of the strong coupling constants $g_{{D D A}}$, $f_{{D^{*} D A}}$ and $h_{{D^{*} D^{*} A}}$}

\begin{figure}
\includegraphics[width=5.9cm,height=6cm]{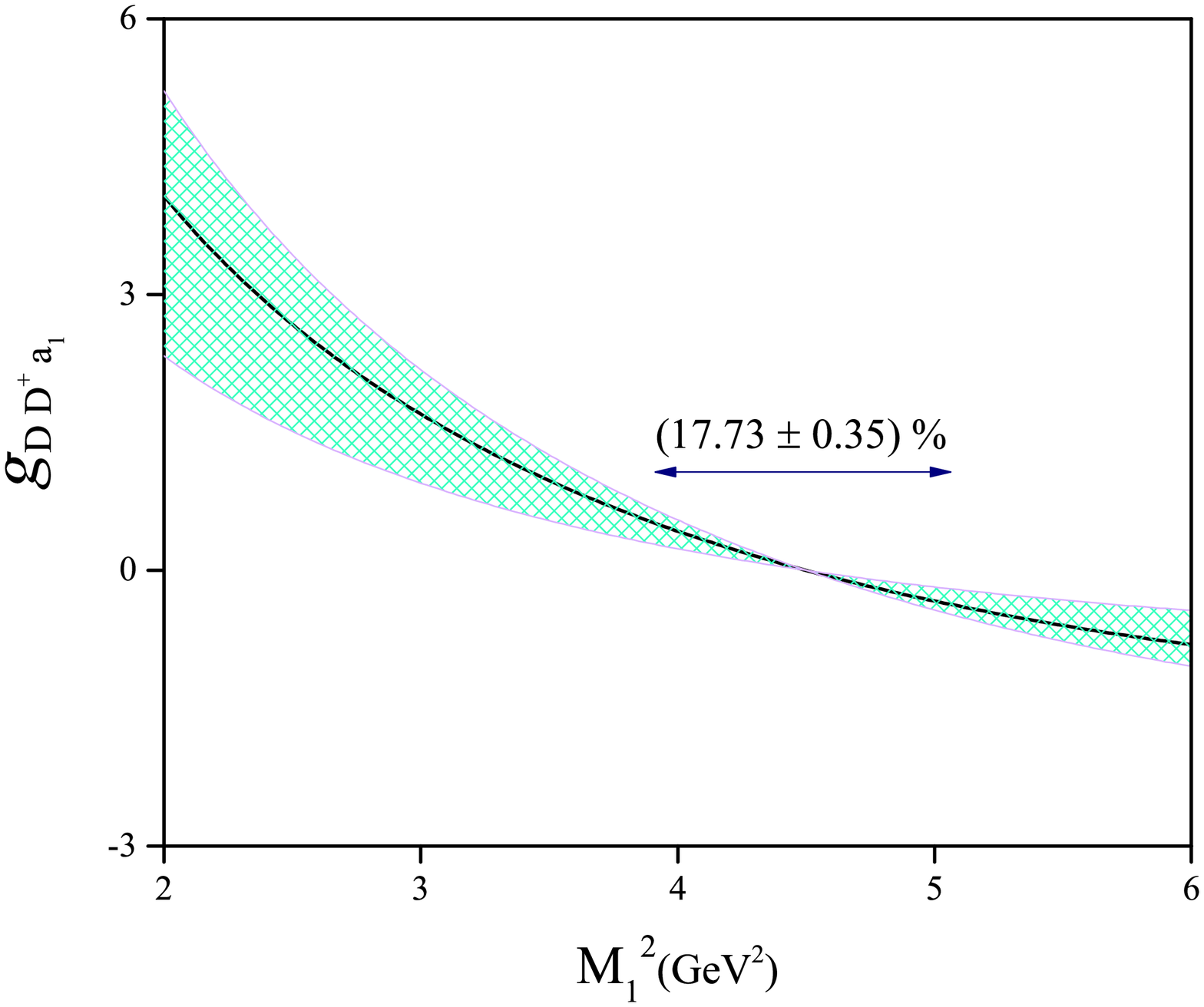}
\includegraphics[width=5.9cm,height=6cm]{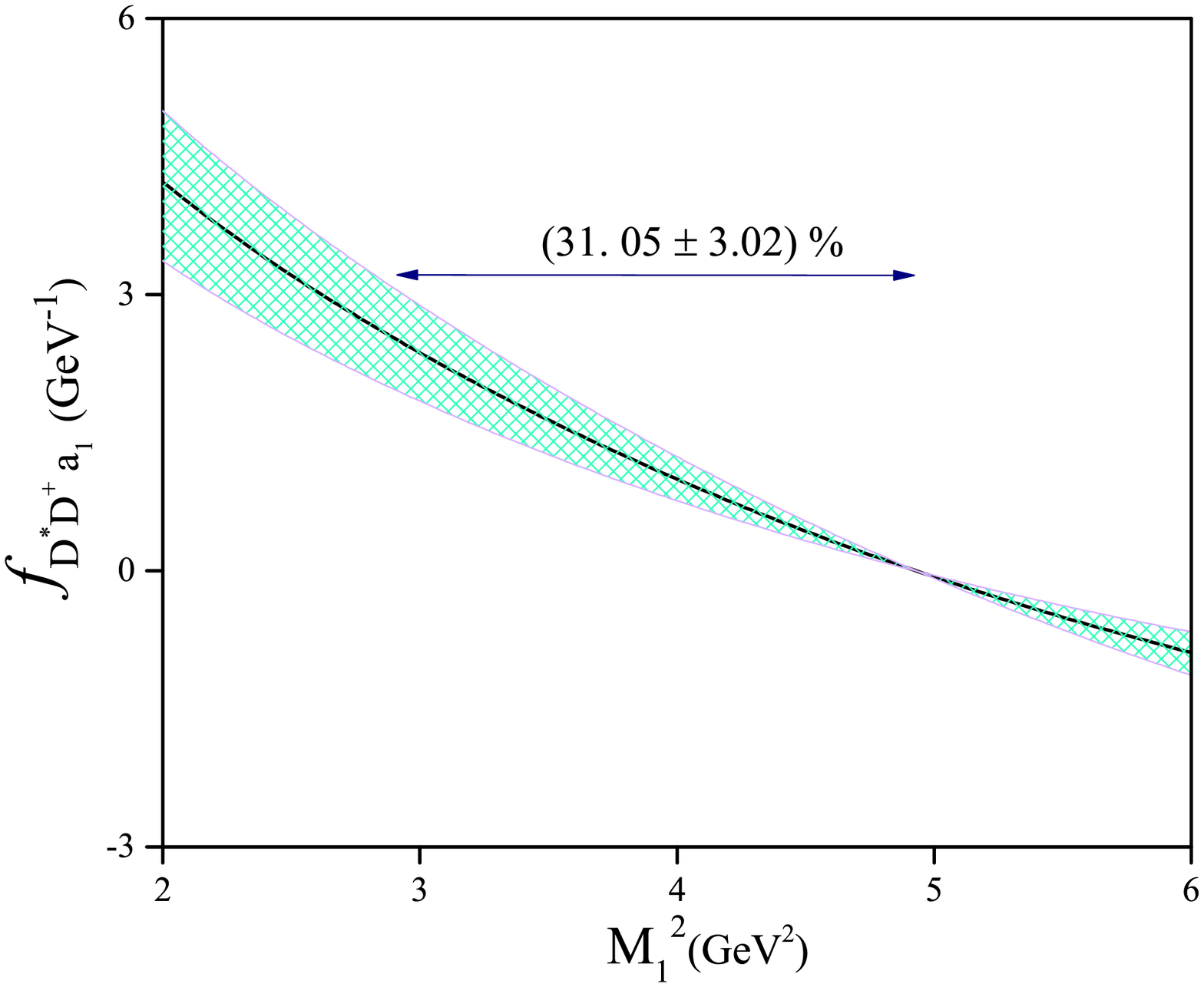}
\includegraphics[width=5.9cm,height=6cm]{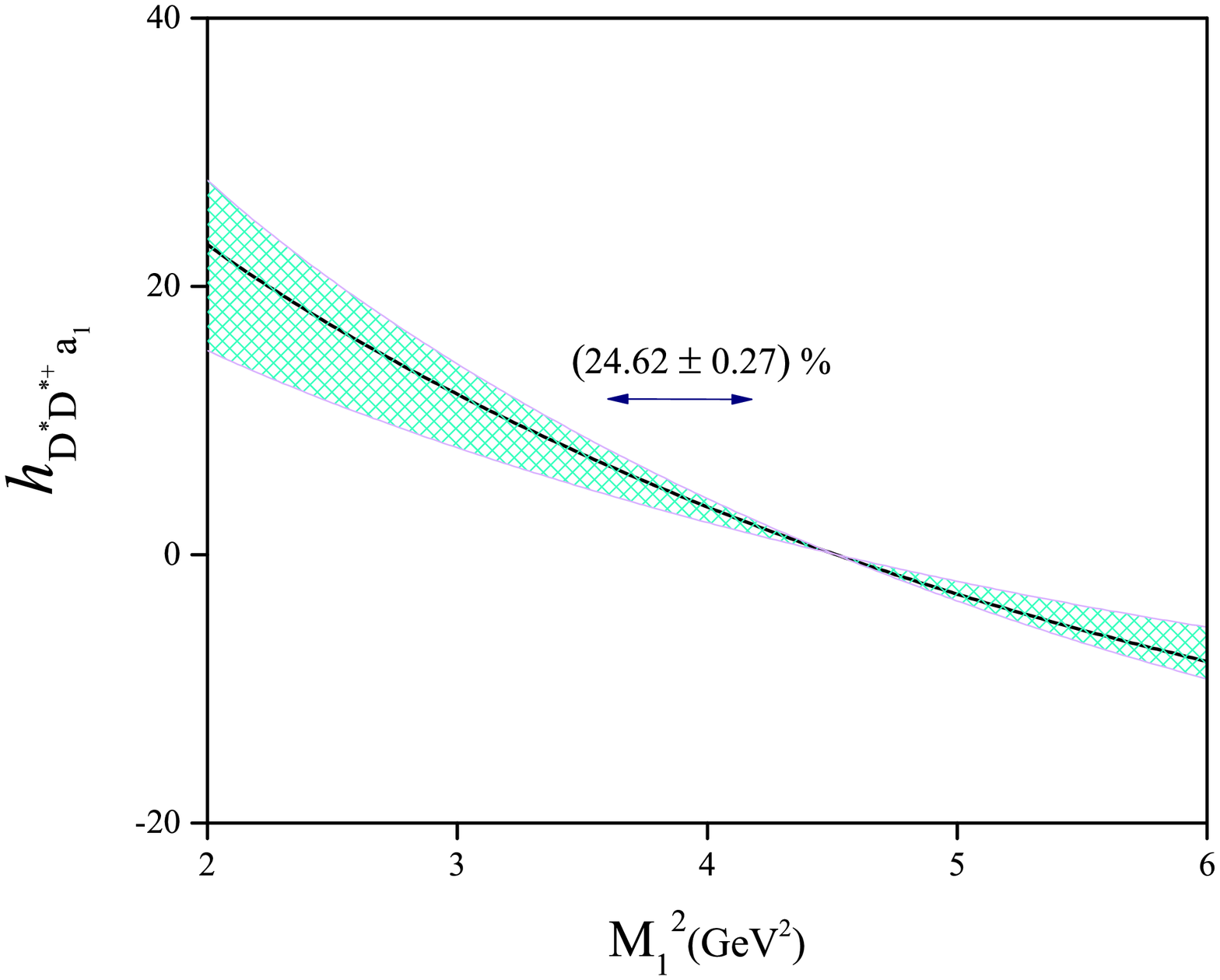}
\caption{ Strong coupling constants $g_{{D D a_1}}$, $f_{{D^{*} D
a_1}}$ and $h_{{D^{*} D^{*} a_1}}$ with their errors as a function
of Borel mass $M_{1}^{2}$. For all the strong coupling constants,
Borel parameter $M_{2}^{2}$  is fixed at $4.5 ~\rm{{GeV}}^{2}$.}
\label{gandfm1}
\end{figure}

\begin{figure}
\includegraphics[width=5.9cm,height=6cm]{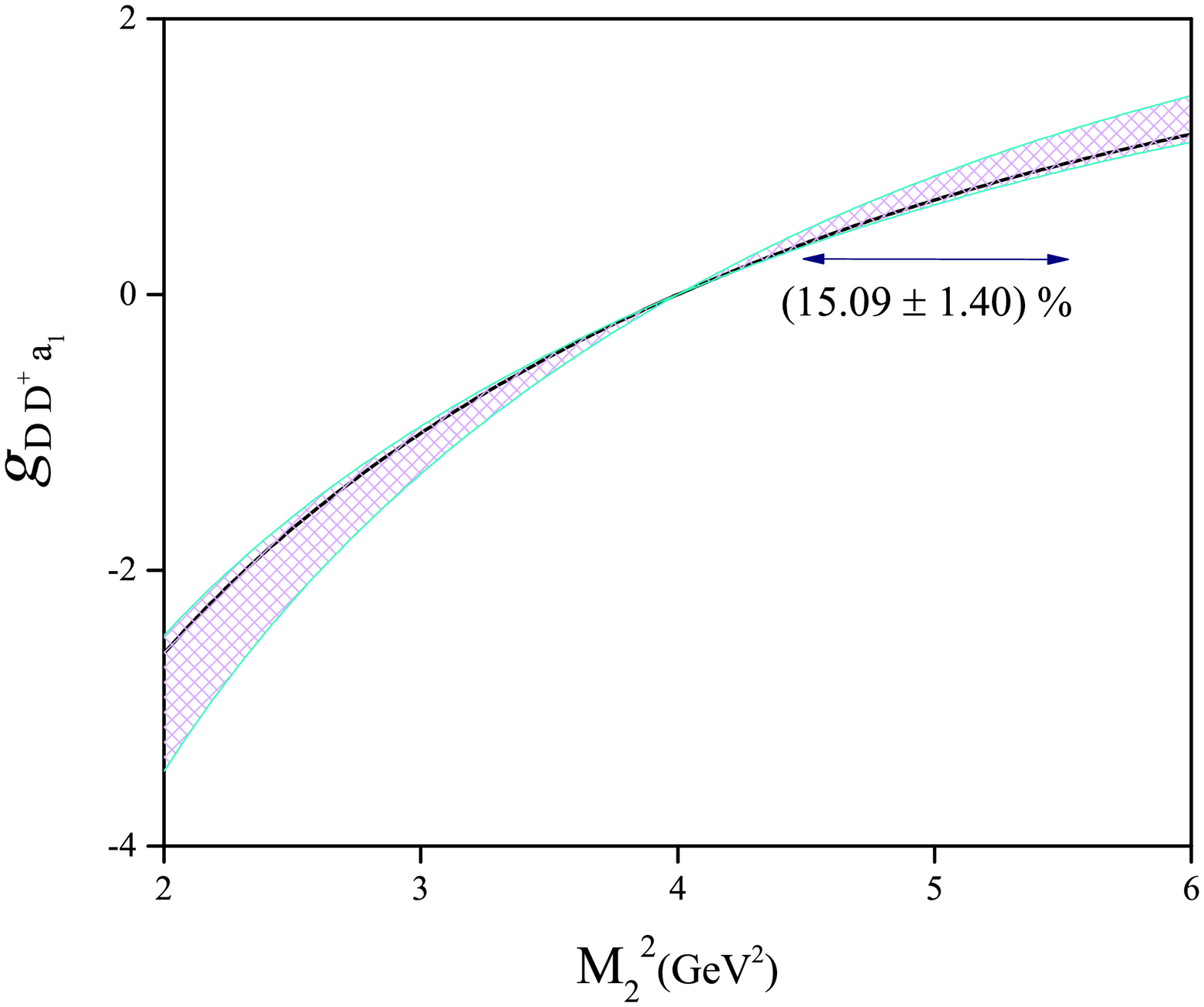}
\includegraphics[width=5.9cm,height=6cm]{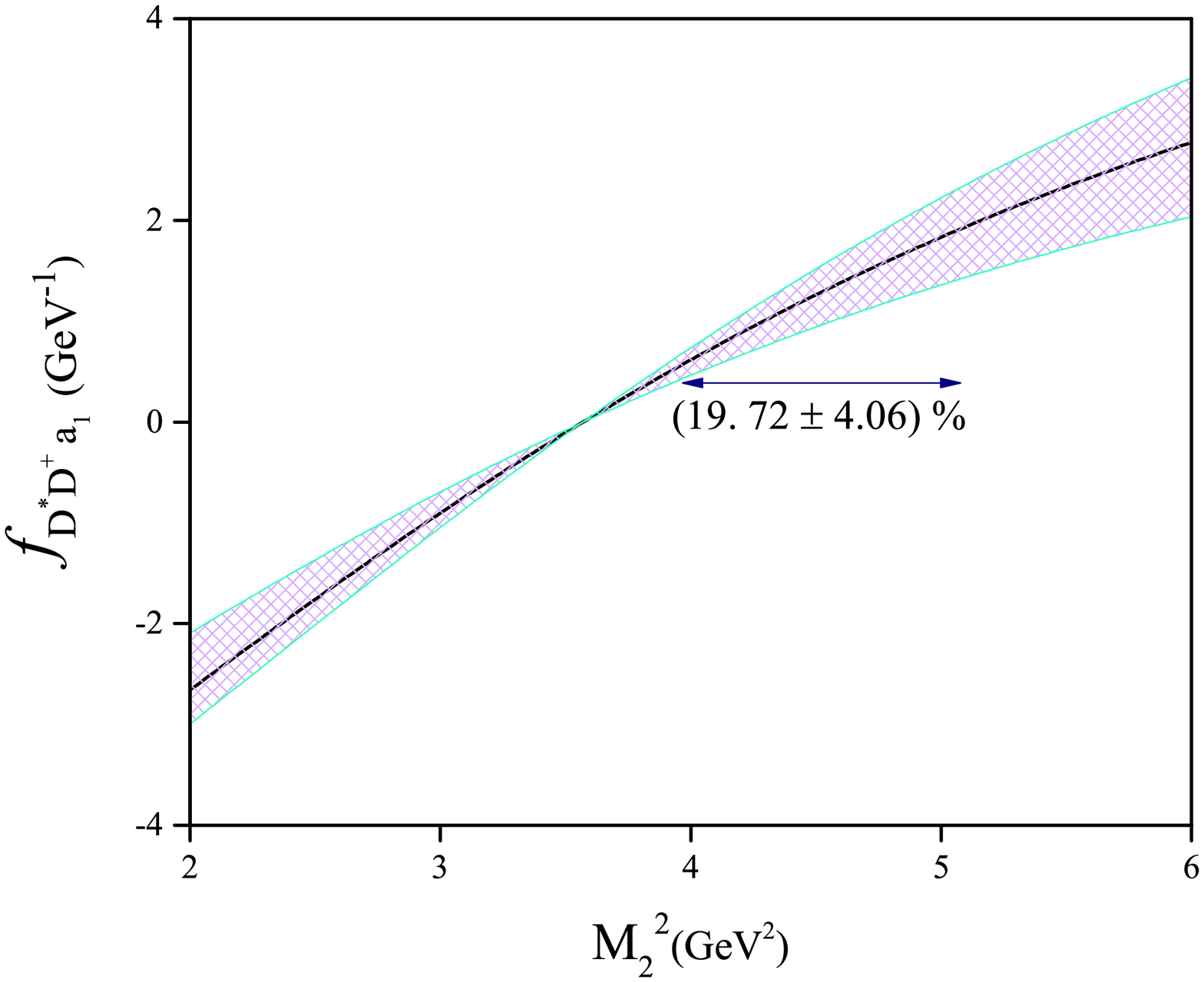}
\includegraphics[width=5.9cm,height=6cm]{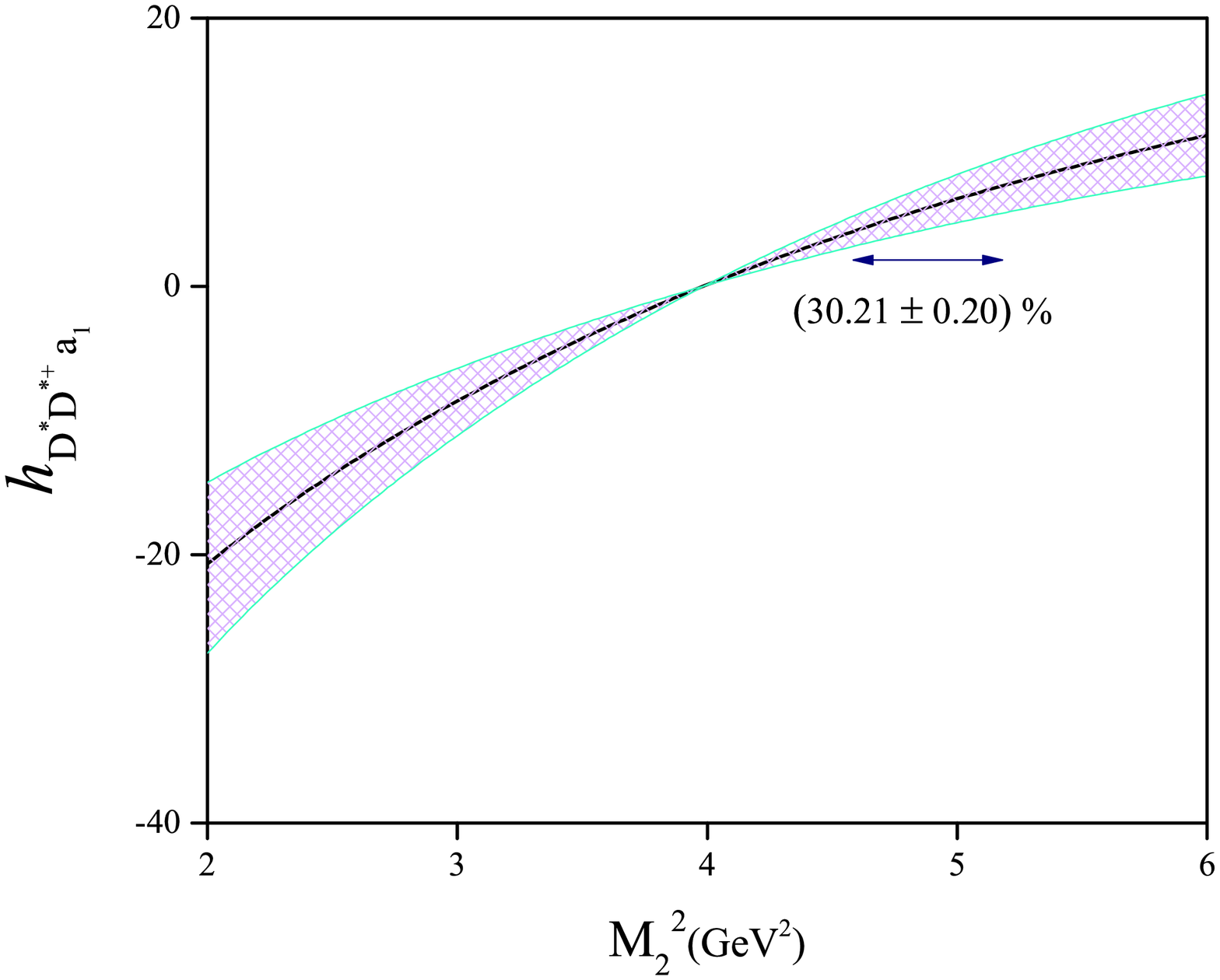}
\caption{The  same as Fig. \ref{gandfm1} but for Borel mass
$M_{2}^{2}$. For all the strong coupling constants the suitable
regions for Borel parameter $M_{1}^{2}$  are shown in Fig
\ref{gandfm1}. } \label{gandfm2}
\end{figure}

Having all the input parameters,  we are ready to do numerical
analysis for the coupling constants. The coupling constants $g_{{D D
A}}$, $f_{{D^{*} D A}}$ and $h_{{D^{*} D^{*} A}}$ in Eq.
(\ref{eq335}) contain  two Borel parameters $M_{1}^{2}$ and
$M_{2}^{2}$. These are not physical quantities, so the coupling
constants should be independent of them. For instance, the
dependence of the strong couplings $g_{{D D a_1}}$, $f_{{D^{*} D
a_1}}$ and $h_{{D^{*} D^{*} a_1}}$  on Borel mass parameters
$M_{1}^{2}$ and $M_{2}^{2}$ are shown in Figs. \ref{gandfm1} and
\ref{gandfm2}, respectively. In these figures, the black lines show
the central values of the coupling constants. The shaded regions are
obtained by using the errors of the input parameters. As can be seen
in Figs. \ref{gandfm1} and \ref{gandfm2}, the strong couplings
$g_{{D D a_1}}$, $f_{{D^{*} D a_1}}$ and $h_{{D^{*} D^{*} a_1}}$ can
be stable within the Borel mass intervals  $4~[4.5]~ \mbox{GeV}^2
<M_1^{2}~[M_2^{2}]<5~[5.5]~ \mbox{GeV}^2$, $3~[4]~ \mbox{GeV}^2
<M_1^{2}~[M_2^{2}]<5~[5]~ \mbox{GeV}^2$, and $3.5~[4.5]~
\mbox{GeV}^2 <M_1^{2}~[M_2^{2}]<4.5~[5.5]~ \mbox{GeV}^2$,
respectively. The percentage of the coupling constant variations at
the suitable intervals of the Borel parameters is displayed in each
plot.
The main uncertainty in $g_{{D D a_1}}$ comes from  $c$ quark mass
($m_c$) and the Gegenbauer moments $a_{0}^{\perp}$, $a_{1}^{\perp}$
and $a_{2}^{\perp}$ of $h_{\|}^{(p)}$ LCDA, while for $f_{{D^{*} D
a_1}}$ and $h_{{D^{*} D^{*} a_1}}$, the main sources of
uncertainties are the $c$ quark mass and $\Phi_{\perp}$ LCDA.

Taking all values and parameters and their uncertainties in Eq.
(\ref{eq335}), the values for the strong coupling constants are
obtained and shown in Table \ref{Tasff}. It is worth mentioning that
the strong coupling constants $g_{{D D a_1}}$ and $h_{{D^{*} D^{*}
a_1}}$ are dimensionless.
\begin{table}[th]
\caption{The values of the strong coupling constants $g_{D D A}$,
$f_{D^* D A}(\mbox{GeV}^{-1})$ and $h_{D^* D^* A}$ obtained in the
LCSR calculation with their uncertainties. } \label{Tasff}
\begin{ruledtabular}
\begin{tabular}{ccccc}
$A$&$a_{1}$&$b_{1}$&$K_{1A}$&$K_{1B}$\\
\hline
$g_{D D A}$                    & $0.38 \pm 0.07$&$1.64 \pm 0.15$&$1.17 \pm 0.49$&$1.51 \pm 0.11$\\
$f_{D^* D A}$ & $1.03 \pm 0.25$&$1.90 \pm 0.72$&$1.36 \pm 0.78$&$2.48 \pm 0.78$\\
$h_{D^* D^* A}$                & $3.67 \pm 1.01$&$6.32 \pm 1.75$&$2.86 \pm 0.95$&$ 6.59 \pm 2.02$\\
\end{tabular}
\end{ruledtabular}
\end{table}

Mesons $a_1$ and $\rho$ have the same quark content, but different
masses and parities, i.e. $\rho$ is a vector ($1^-$) and $a_1$ is a
axial vector ($1^+$). The values of the strong couplings $g_{D D
\rho}$, $f_{D^{*} D \rho}$ and $h_{D^*D^*\rho}$ are evaluated as
$(1.31 \pm 0.29)$, $(0.89 \pm 0.15)\,\mbox{GeV}^{-1}$ and $(6.6 \pm
0.31)$ using the SR method in Refs. \cite{GangWang,Bracco2008}.
According to Table \ref{Tasff}, only the strong couplings $f_{D^{*}
D a_1}$ and $f_{D^{*} D \rho}$ are approximately equal.

\begin{figure}
\includegraphics[width=5.9cm,height=6cm]{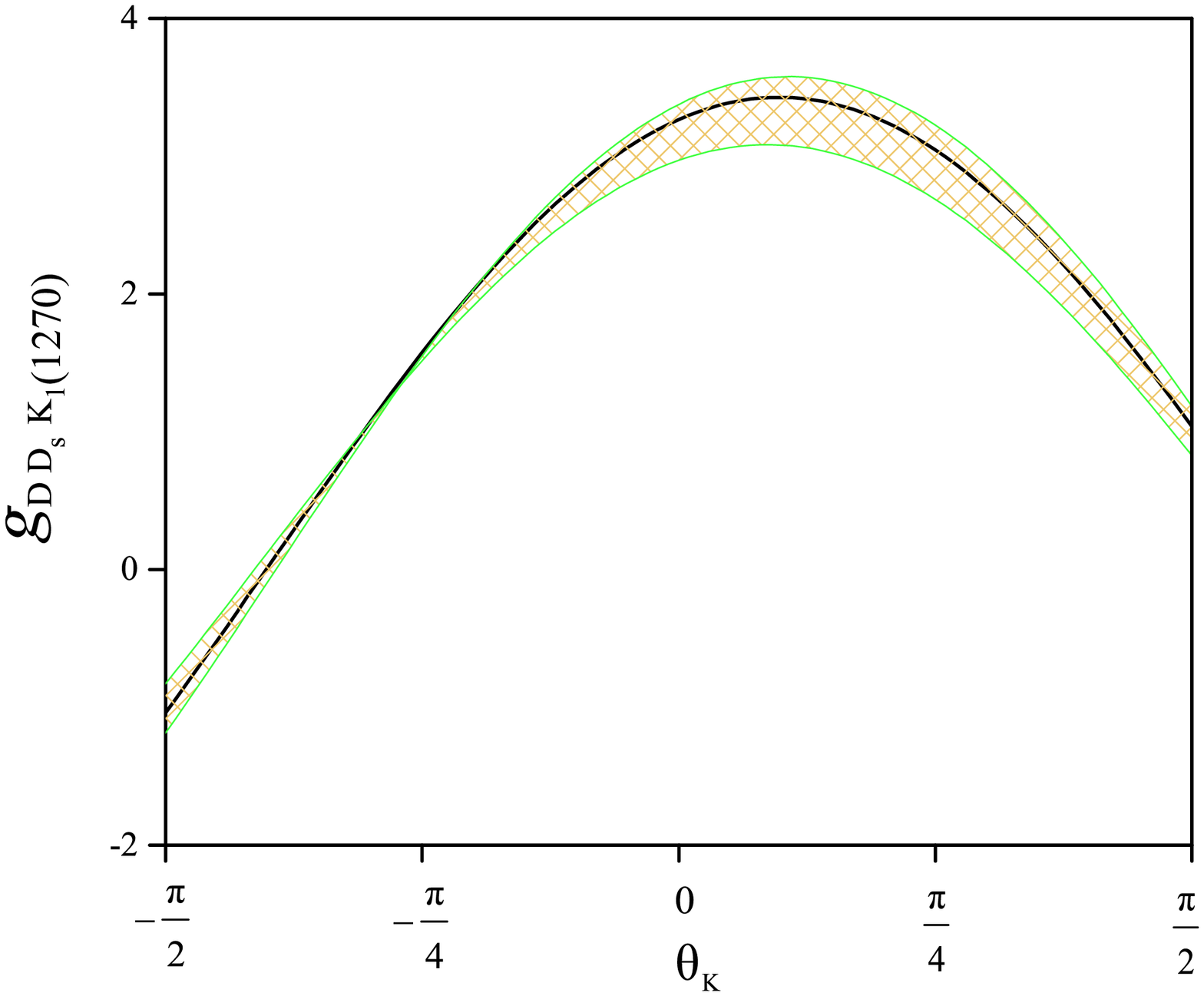}
\includegraphics[width=5.9cm,height=6cm]{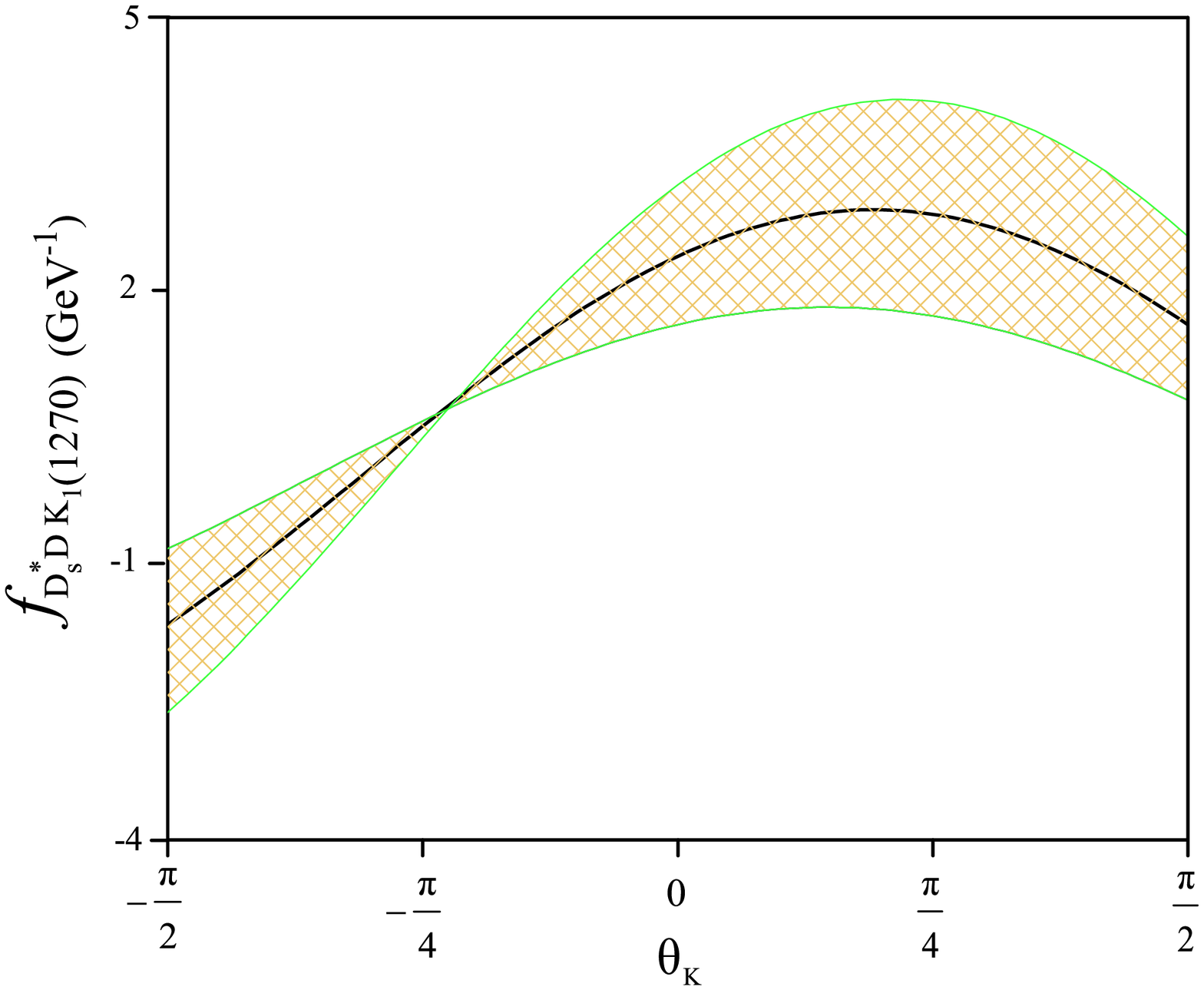}
\includegraphics[width=5.9cm,height=6cm]{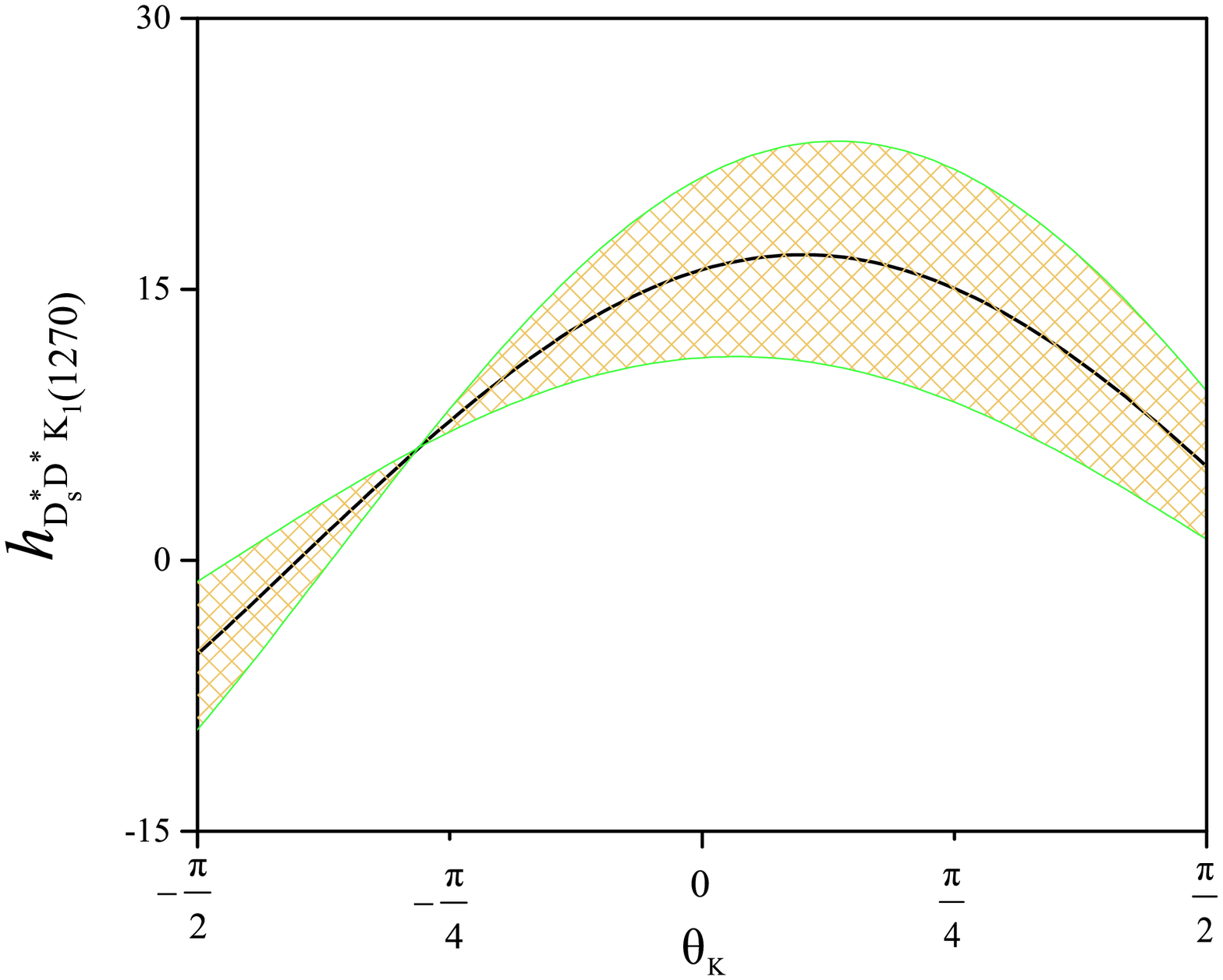}
\includegraphics[width=5.9cm,height=6cm]{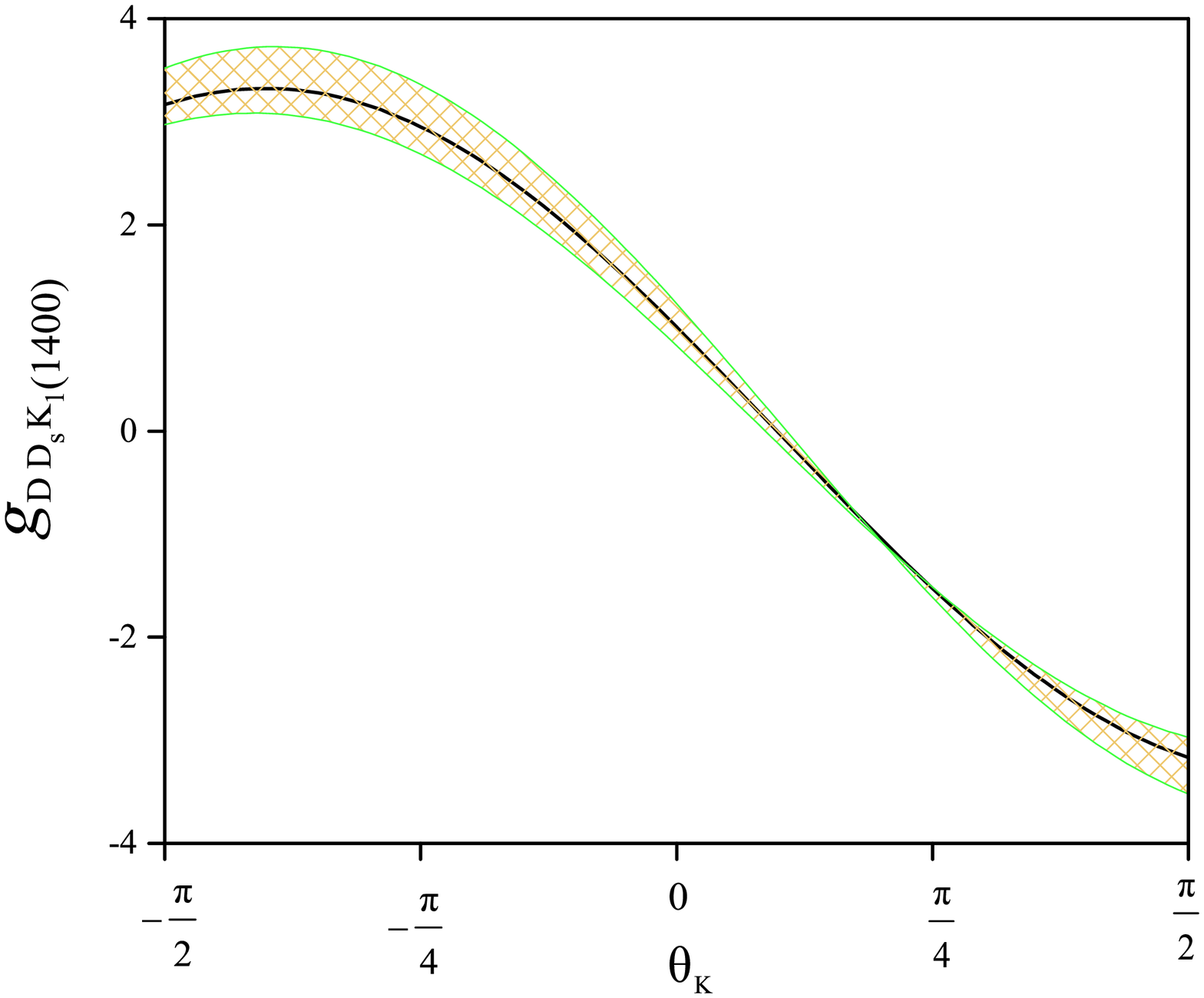}
\includegraphics[width=5.9cm,height=6cm]{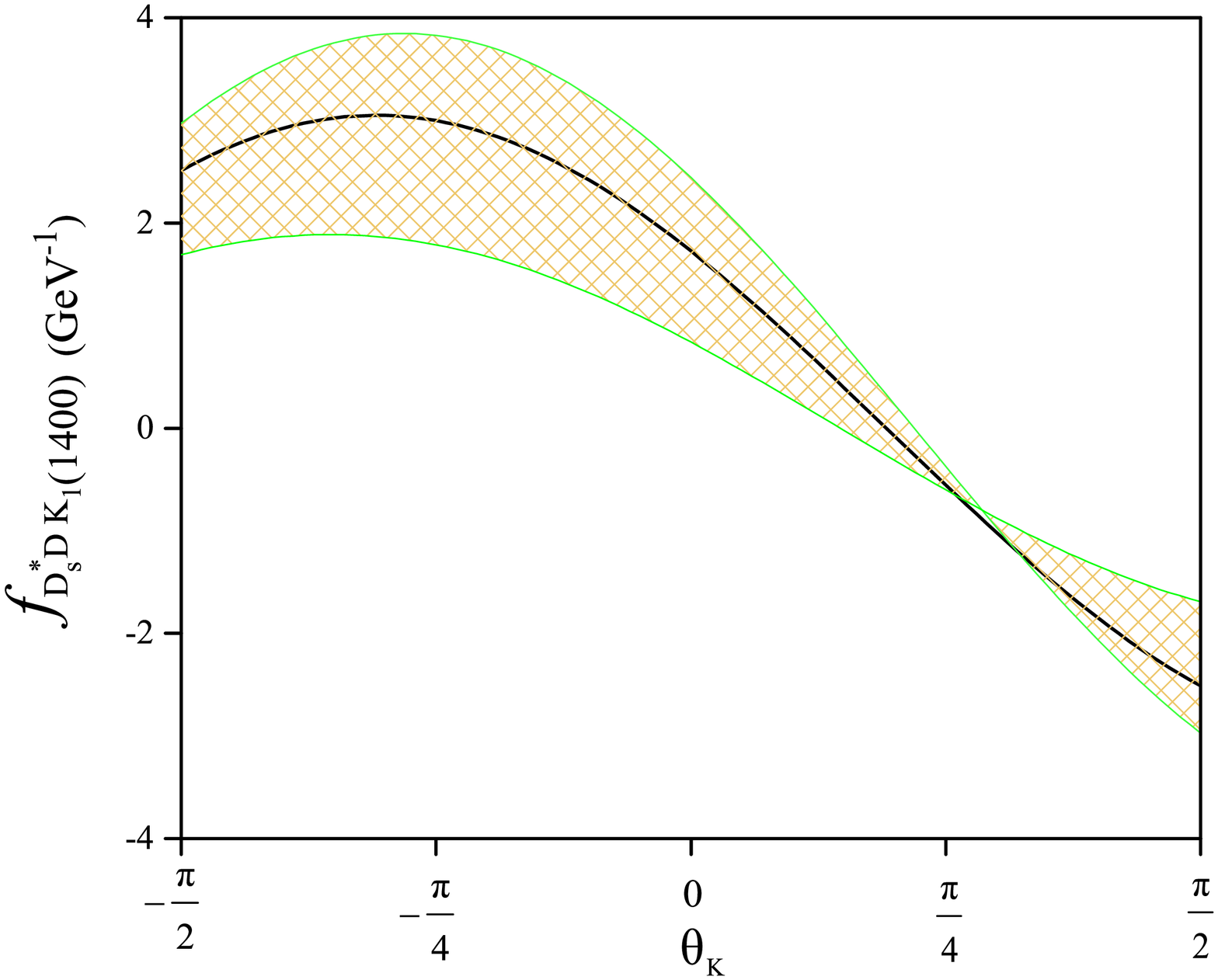}
\includegraphics[width=5.9cm,height=6cm]{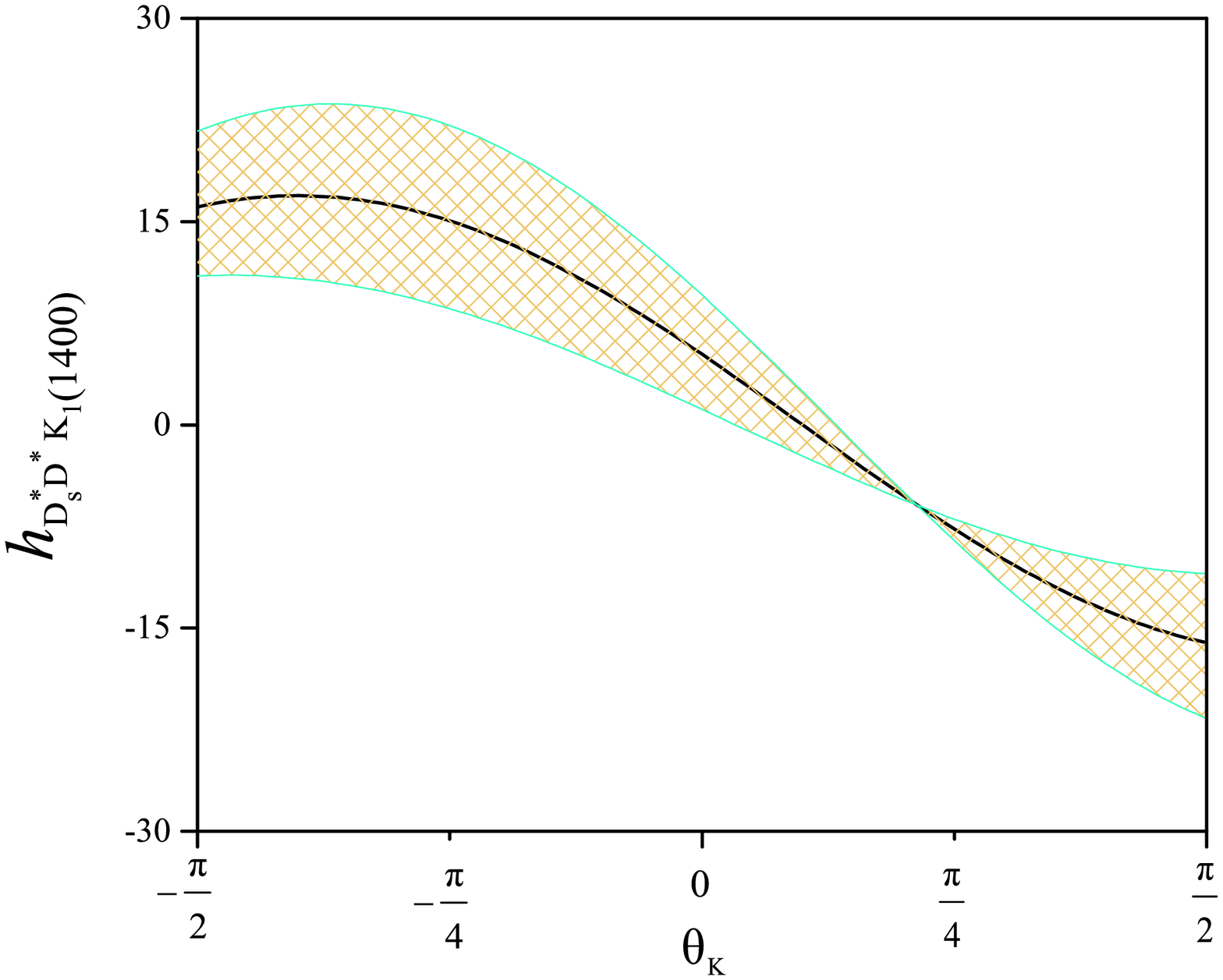}
\caption{The strong coupling constants $g_{D_s D K_1}$, $f_{D_s^* D
K_1}$ and $h_{D_s^* D^* K_1}$ for $K_1(1270)$ and $K_1(1400)$ as a
function of the mixing angle $\theta_{K}$. } \label{fandgk1}
\end{figure}

The strong coupling constants $g_{D_s D K_1}$, $f_{D_s^* D K_1}$ and
$h_{D_s^* D^* K_1}$ for $K_1(1270)$ and $K_1(1400)$ are plotted in
Fig. \ref{fandgk1}, as a function of the mixing angle $\theta_{K}$.
The uncertainty regions are also displayed  in this figure.

The values of the strong couplings $g_{D_s D K_1}$, $f_{D_s^* D
K_1}$ and $h_{D_s^* D^* K_1}$  depend on the mixing angle
$\theta_K$.  The mixing angle $\theta_K$  can be determined by the
experimental data \cite{Burakovsky,Suzuki,HYCheng,Hatanaka2}. A new
research for the value of $\theta_K$ indicates that this mixing
angle is around either $\pm 33^{\circ}$ or $\pm 58^{\circ}$
\cite{TaKoYa,DiOlGi}. On the other hand, the recent experimental
values for the branching ratios of the $B \to K_1(1270, 1400) \pi$
decays are reported at $\theta_{K}=(72 \pm 3)^{\circ}$ by BABAR
\cite{Aubert2010}. For the next calculations and comparisons, we
need to the values of the aforementioned coupling constants at
$\theta_{K}=-58^{\circ}$, $-37^{\circ}$, $-33^{\circ}$, $32^{\circ}$
and $72^{\circ}$. Therefore,  these values are presented in Table
\ref{Ta2}.
\begin{table}[th]
\caption{ Values of the strong coupling constants $g_{D_s D K_1}$,
$f_{D_s^* D K_1}\,(\mbox{GeV}^{-1})$, and $h_{D_s^* D^* K_1}$ at the
various mixing angles.} \label{Ta2}
\begin{ruledtabular}
\begin{tabular}{cccccc}
$\theta_{K}$ &$-58^{\circ}$&$-37^{\circ}$&$-33^{\circ}$&$32^{\circ}$&$72^{\circ}$\\
\hline
$g_{{D_{s} D K_{1}(1270)}}$ &$0.87\pm 0.05$&$1.94\pm 0.20$&$1.95 \pm 0.25 $&$3.31\pm 0.26$&$1.98\pm 0.19$\\
$f_{{D_{s}^{*} D K_{1}(1270)}}$ &$-0.10\pm 0.35$&$0.88\pm 0.08$&$2.39 \pm 1.12$&$2.98\pm 1.16$&$2.29\pm 1.01$\\
$h_{{D_{s}^{*} D^{*} K_{1}(1270)}}$ &$2.23\pm 0.79$&$9.52\pm 1.38$&$10.20 \pm 5.58$&$16.37\pm 6.28$&$9.91\pm 5.35$\\
\hline
$g_{{D_{s} D K_{1}(1400)}}$ &$3.21\pm 0.26$&$2.73\pm 0.38$&$-2.59 \pm 0.16$&$-0.84\pm 0.05$&$-2.69\pm 0.13$\\
$f_{{D_{s}^{*} D K_{1}(1400)}}$ &$2.89\pm 0.81$&$2.75\pm 0.88$&$-1.71 \pm 0.35$&$0.10\pm 0.30$&$-1.76\pm 0.40$\\
$h_{{D_{s}^{*} D^{*} K_{1}(1400)}}$ &$16.37\pm 7.05$&$7.63\pm 5.02$&$-13.68 \pm 3.49$&$-4.23\pm 0.69$&$-13.69\pm 3.02$\\
\end{tabular}
\end{ruledtabular}
\end{table}

\subsection{ Branching ratio analysis  of the non-leptonic  $B^0\to {K}_{1}^{+} \pi^-$ decay}
In this section, we want to evaluate the branching ratio values for
the non-leptonic $B^0\to {K}_{1}^{+}(1270, 1400) \pi^-$ decays.
According to Refs. \cite{Isola1, Isola2}, the amplitude of $B^0\to
{K}_{1}^{+} \pi^-$ decay, $\mathcal{M}_{{K_1 \pi}}$, is written in
two parts; the short-distance contribution (SD) and the
long-distance ones (LD), as:
\begin{eqnarray}\label{eqamp}
{\mathcal{M}}_{K_1  \pi} = {\mathcal{M}}_{SD} + {\mathcal{M}}_{LD}.
\end{eqnarray}
In the above phrase, $ {\mathcal{M}}_{SD}$ is written using the
effective Hamiltonian for the non-leptonic $B$ decays in the
factorization approximation as \cite{Fleischer, Fleischer1,
Fleischer2, Fleischer3, Fleischer4}:
\begin{eqnarray}\label{eqsdam}
{\mathcal{M}}_{SD}(B^0\to {K}_{1}^{+} \pi^-)=\, G_F\,\sqrt{2} \,
F_1^{B\to\pi}(m_{K_{1}}^2)\,f_{K_{1}}\, m_{K_{1}}\; \left [
\frac{}{} V_{ub}^\ast V_{us} \;  a_2\, -\, V_{tb}^\ast V_{ts}\;
\left(a_4+a_{10}\right ) \right ]\left( \varepsilon\cdot p_B\right
),
\end{eqnarray}
where $G_F$ is the Fermi constant, $V_{ij}$ is the CKM matrix
element, $\varepsilon$ is the polarization vector of $K_1$ meson and
$ a_i = C_i + \frac{C_{i-1}}{3}$ ($C_i$ is the Wilson coefficient).
In addition $F_1^{B\to\pi} (m_{K_{1}}^2)$ is the transition form
factor of the semileptonic $B \to \pi$ decay estimated in
$m_{K_{1}}^2$ \cite{Casalbuoni1997}.

\begin{figure}
\includegraphics[width=10cm,height=5cm]{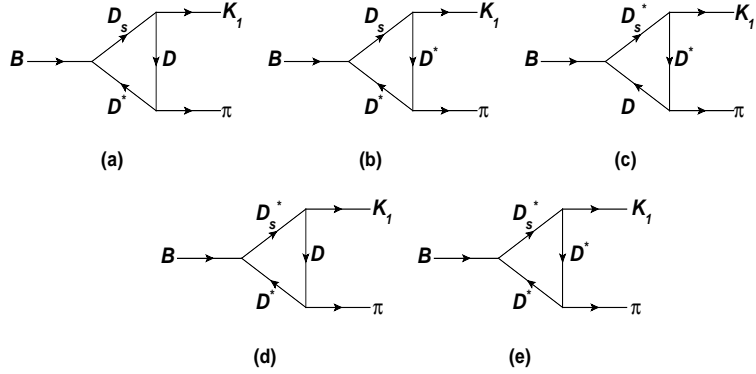}
\caption{Diagrams for the $B\to K_1\pi$ decay with $D^{(*)}_s$ and
$D^{(*)}$ intermediate states. } \label{F31}
\end{figure}

For the long-distance part of the amplitude, diagrams displayed in
Fig. \ref{F31} are considered.
As shown in this figure, $D_s^{(*)}$ and $D^{(*)}$  mesons are
considered as the intermediate states in the decay process of $B^0$
to ${K}_{1}^{+} \pi^-$; first, $B$ meson decays into a $D^{(*)}_s
D^{(*)}$ intermediate states, and then these two particles produce
the final mesons $K_{1}$ and $\pi$ by exchanging a $D^{(*)}$ meson.
In this view, the ${\mathcal{M}}_{LD}$ is given by the following
formula:
\begin{eqnarray}\label{eqld}
{\mathcal{M}}_{LD}=\mathrm{Re}({\mathcal{M}}_{LD})+\mathrm{Im}({\mathcal{M}}_{LD}).
\end{eqnarray}
Using the charming penguin diagrams in Fig. \ref{F31},  the
imaginary part of ${\mathcal{M}}_{LD}$ can be computed as
\begin{eqnarray}\label{eqimld}
\mathrm{Im}({\mathcal{M}}_{LD})& =& \frac{m_D}{32 \pi^2 m_B}
\sqrt{\omega^{*2}-1} \int d{\bf n}~ {\cal M}\left[B(v)\to
D_s^{(*)}(q)D^{(*)}(v')\right]\, {\cal M}\left[D_s^{(*)}D^{(*)}\to
K_{1} \pi\right],
\end{eqnarray}
where the integration is over the solid angle. The amplitude for
$B\to D_s^{(*)}D^{(*)}$ transition is computed as
\begin{eqnarray}
{\cal M}\left[B(v)\to D_s(q) D^{*}(\varepsilon^*,
v^\prime)\right]&=& - K
\,(m_B+m_{D_s})\,\varepsilon\cdot v\,,\\
{\cal M}\left[B(v)\to D_s^{*}(\varepsilon ,q) D(v^\prime)\right]&=&
- K\,m_{D_s^*}\,
\varepsilon\cdot(v+v^\prime)\,, \\
{\cal M}\left[B(v)\to D_s^{*}( \varepsilon, \, q )D^{*}(
\varepsilon^*, v^\prime )\right]&=& -\,i  K\,
m_{D_s^*}\varepsilon^{\mu}\varepsilon^{*\alpha}\left(i\epsilon_{\alpha\lambda\mu\sigma}
v^{\prime\lambda}v^\sigma-g_{\mu\alpha}(1+\omega^*)+v_\alpha
v_{\mu}^\prime \right)\ .
\end{eqnarray}
In these phrases $K$ and $\omega^*$ are as follows:
\begin{eqnarray}
K&=&\frac{\sqrt{2}\, G_{F}\,a_2 }{1+\omega^*}\,V_{cb}^* V_{cs}\
\sqrt{m_B m_{D^*}} f_{D^{(*)}_s},~~~~~~ \omega^*=
\frac{m_B^2+m_{D^{(*)}}^2-m^2_{D^{(*)}_s}}{2 m_{D^{(*)}} m_B}.
\end{eqnarray}
It should be recalled that the factorization with the following
kinematics are used to compute ${\cal M}[B\to D_s^{(*)}D^{(*)}]$:
\begin{eqnarray}
p^\mu=m_B v^\mu=(m_B,\vec 0 )\ ,\ ~~~~
p'^\mu=m_{D^{(*)}}v^{\prime\mu}\ ,\ ~~~~ q=p-p'.
\end{eqnarray}

On the other hand, the heavy quark effective lagrangian is used to
estimate ${\cal M}[D_s^{(*)}D^{(*)}\to K_{1} \pi]$.  The result for
the sum of diagrams (a) and (b) in  Fig. \ref{F31} is obtained as
\begin{eqnarray}\label{eqsab}
{\cal M}^{(a+b)}\left[D_s(q)\, D^{*}(\varepsilon^*, v^\prime)\to
K_{1}(p_{K_{1}},\varepsilon)\pi(p_\pi)\right]&=&-\frac{\sqrt 2\, g\,
F^2(|\vec p_\pi|) }{f_\pi}\,g_V \sqrt{
\frac{m_{D^*}}{m_{D_s}}}\,\varepsilon^*_\eta\,\varepsilon_\sigma\nonumber\\
&&\times\left\{\frac{2\beta_{1}\,m_{D_s}\, q^\sigma p_\pi^\eta} {
(m_{D_s}v^\prime-p_\pi)^2-m^2_{D_s} }\ +\
\frac{4\,\lambda\,m_{D^*}\,G^{\sigma\eta}(p_\pi,p_{K_{1}},
v^\prime)}{ (m_{D^*}v^\prime-p_\pi)^2-m^2_{D^*} }\right\},
\end{eqnarray}
where $F(|\vec p_\pi|)\, = \,0.065$ \cite{Isola2}, $g_V\simeq 5.8$
\cite{Bando}, $g = 0.59 \pm 0.07 \pm 0.01$ \cite{Ahmed}. For diagram
(c),  amplitude is obtained as:
\begin{eqnarray}\label{eqsc}
&&{\cal M}^{(c)}\left[D_s^{*}(\varepsilon^* ,q) D(v^\prime)\to
K_{1}(p_{K_{1}},\varepsilon)\pi(p_\pi)\right]= -\frac{\sqrt 2\,
g\,m_{D^*}\, F^2(|\vec p_\pi|) }{f_\pi} g_V \sqrt{
\frac{m_{D}}{m_{D^*_s}}}\,\frac{\varepsilon^*_\eta\varepsilon_\sigma}{
(m_{D^*_s}v^\prime-p_\pi)^2-m^2_{D^*_s} }\nonumber\\
&&\times\left\{2\beta_{1} q^\sigma\left(p_\pi^\eta\,
-\,\frac{v^\prime\cdot p_\pi}{m_{D^*_s}} \,p_{K_{1}}^\eta
\right)-4\,\lambda\, m_{D^*_s}H^{\sigma\eta}(p_\pi,p_{K_{1}},
v^\prime) \right\},
\end{eqnarray}
and finally,  the sum of diagrams (d) and (e) leads to:
\begin{eqnarray}\label{eqsde}
&&{\cal M}^{(d+e)}\left[D_s^{*}(\varepsilon^* ,q)
D^{*}(\hat\varepsilon^*, v^\prime) \to
K_{1}(p_{K_{1}},\varepsilon)\pi(p_\pi)\right]= \frac{\sqrt 2\,
g\,m_{D^*}\, F^2(|\vec p_\pi|) }{f_\pi}\,g_V \sqrt{
\frac{m_{D^*}}{m_{D^*_s}}}\,\epsilon^{\alpha\mu\nu\eta}\,\varepsilon^*_\tau
\varepsilon_\sigma\hat\varepsilon^*_\rho \cr &&\cr &&\times\Big\{
\frac{q_\alpha
(p_{K_{1}})_\mu\,\delta^\sigma_\nu\delta^\tau_\eta}{(m_{D_s^*}v^\prime-p_\pi)^2-m^2_{D_s^*}}
\frac{4\lambda \,m_{D_s^*}\,p_\pi^\rho }{m_{D^*}}\,+\,
\frac{v^\prime_\alpha (p_\pi)_\mu\delta^\rho_\nu
}{(m_{D^*}v^\prime-p_\pi)^2-m^2_{D^*}}\left( 2\beta_{2}\,
q^\sigma\delta^\tau_\eta\,+\,4\lambda\,
m_{D^*_s}[p_{K_{1}}^\tau\delta^\sigma_\eta-(p_{K_{1}})_\eta
g^{\sigma\tau}]\right) \Big\}.
\end{eqnarray}
In these phrases $G^{\sigma\eta}$ and $H^{\sigma\eta}$ are
\begin{eqnarray}
G^{\sigma\eta}(p_\pi,p_{K_{1}}, v^\prime)&=&-(v^\prime\cdot q)
(g^{\sigma\eta}[p_{K_{1}}\cdot p_\pi] -p_\pi^\sigma p_{K_{1}}^\eta)
-(q\cdot
p_\pi)(v^{\prime\sigma}p_{K_{1}}^\eta-g^{\sigma\eta}[v^\prime\cdot
p_{K_{1}}])\nonumber\\ &-& q^\eta (p_{\pi}^{\sigma} [p_{K_{1}}\cdot
v^\prime] - v^{\prime\sigma} [p_{K_{1}}\cdot p_\pi])\,,
\nonumber\\
H^{\sigma\eta}(p_\pi,p_{K_{1}}, v^\prime)&=& g^{\sigma\eta}
(p_{K_{1}}\cdot p_\pi\,-\, \frac{v^\prime\cdot
p_\pi}{m_{D_s^*}}[m_{K_{1}}^2-p_{K_{1}}\cdot
q])-p_{K_{1}}^\eta(p_\pi^\sigma\, +\,\frac{v^\prime\cdot
p_\pi}{m_{D_s^*}} \,q^\sigma)\,.
\end{eqnarray}
Moreover, the parameters $\beta$ and $\lambda$ are related to the
strong coupling constants $g_{D_s D K_{1}}$, $f_{D_s^{*} D K_{1}}$
and $h_{D_s^{*}D^{*} K_{1}}$ as
\begin{eqnarray}\label{eqbetal}
\beta_{1}= \frac{\sqrt{2}\, g_{D_s D K_{1}}}{2\,g_V},~~~~~~~
\lambda=\frac{\sqrt{2}\, f_{D_s^{*} D K_{1}}}{2\,g_V},~~~~~~~
\beta_{2}=\frac{\sqrt{2}\, h_{D_s^* D^* K_{1}}}{2\,g_V}.
\end{eqnarray}
Similar method of the imaginary part is used to calculate the real
part of the LD amplitude ($\mathrm{Re}({\mathcal{M}}_{LD})$).

\begin{figure}
\includegraphics[width=7cm,height=7cm]{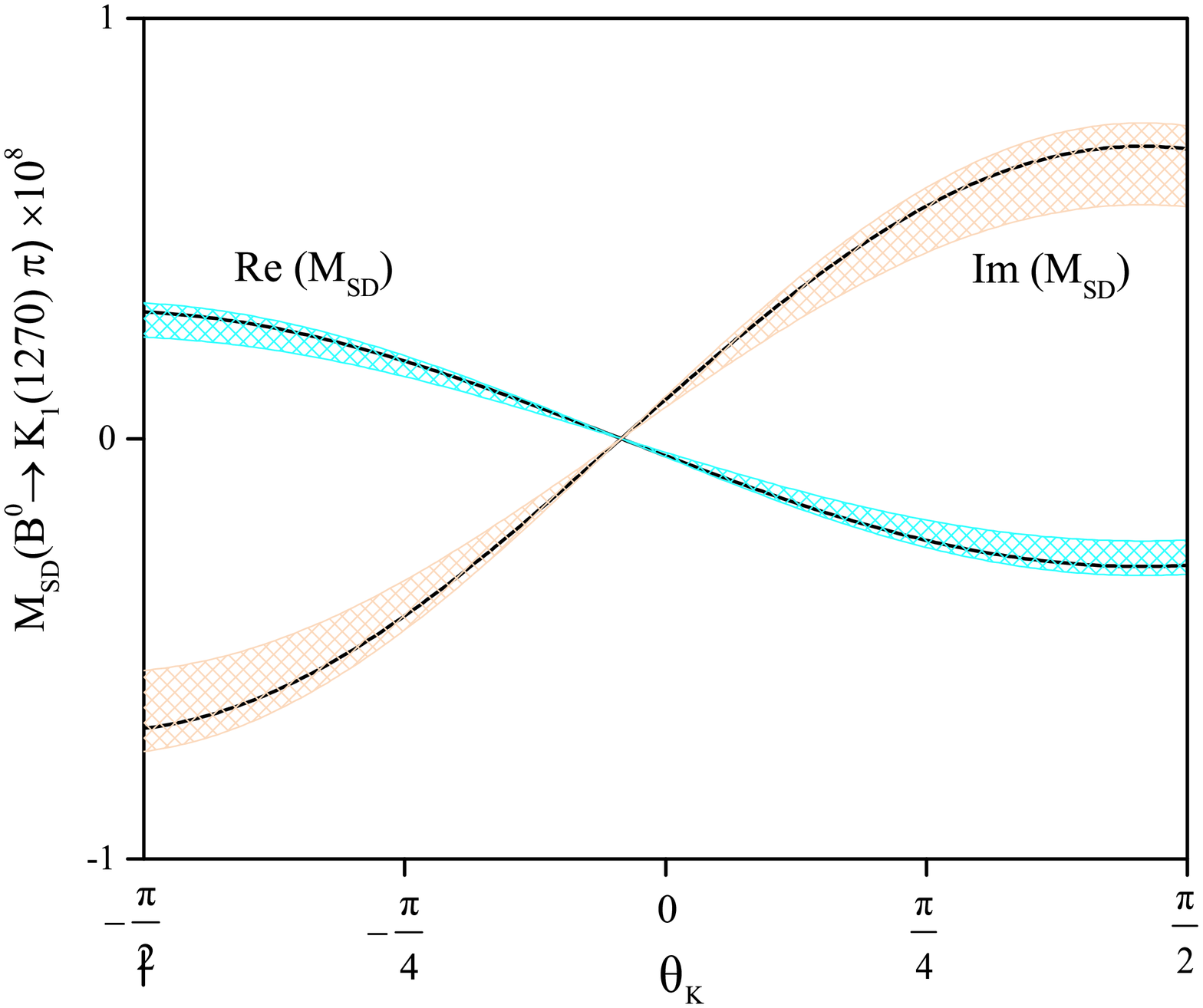}
\includegraphics[width=7cm,height=7cm]{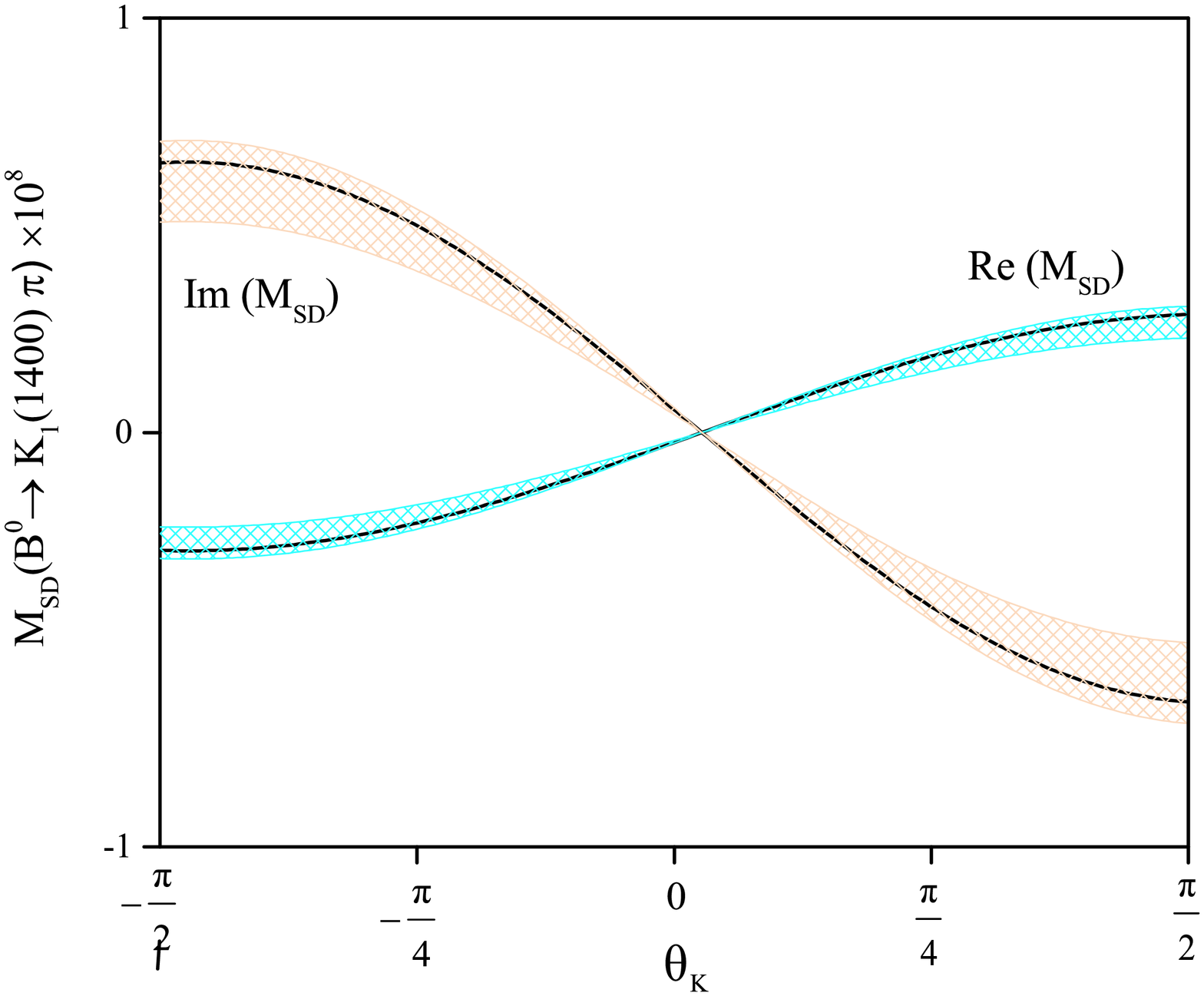}
\caption{ The real and imaginary parts of $\mathcal{M}_{SD}$ as a
function of the mixing angle $\theta_{K}$ for  $B^0\to
{K}_{1}^{+}(1270) \pi^-$ and $B^0\to {K}_{1}^{+}(1400) \pi^-$
decays.} \label{sdamp}
\end{figure}

\begin{figure}
\includegraphics[width=7cm,height=6cm]{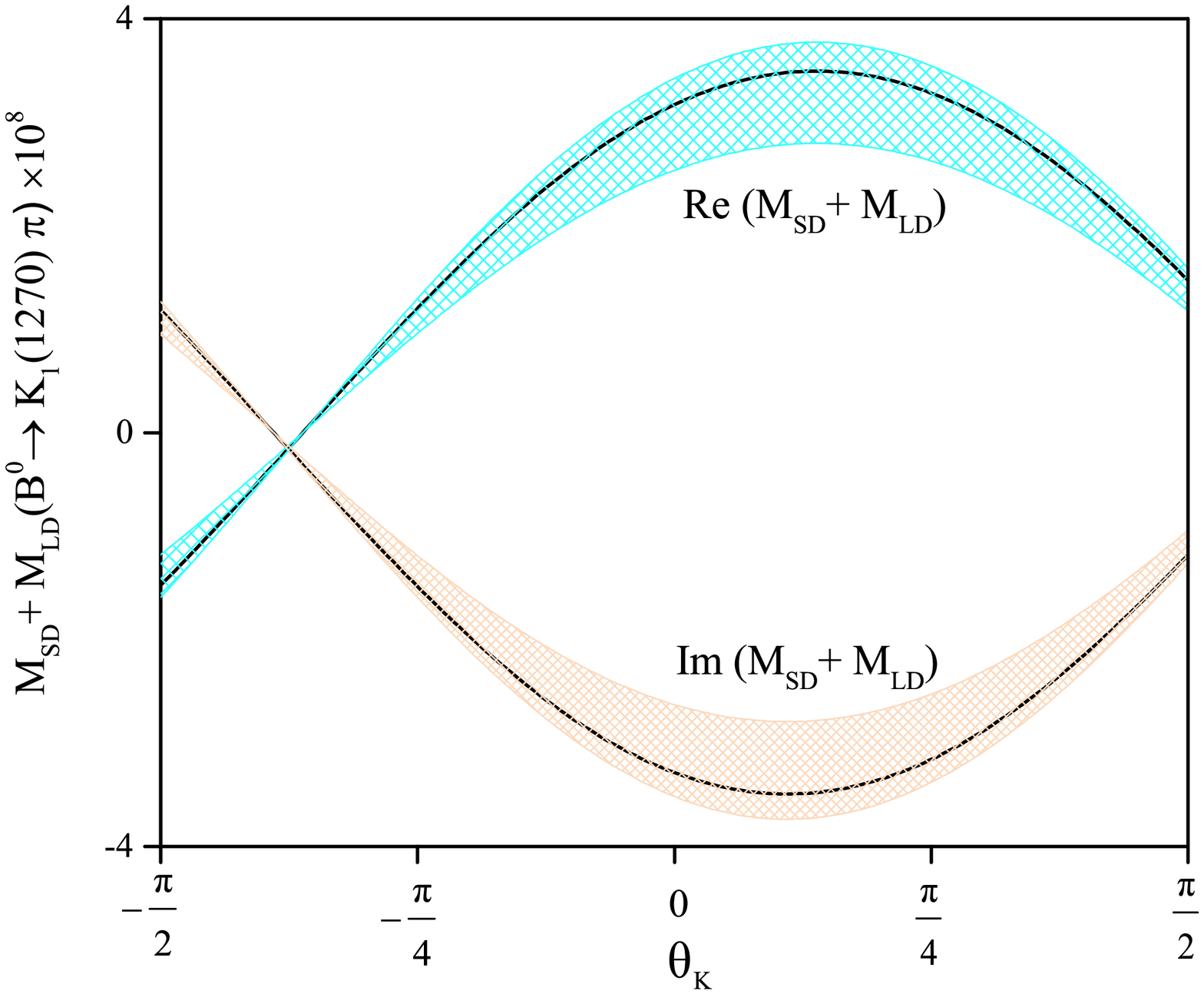}
\includegraphics[width=7cm,height=6cm]{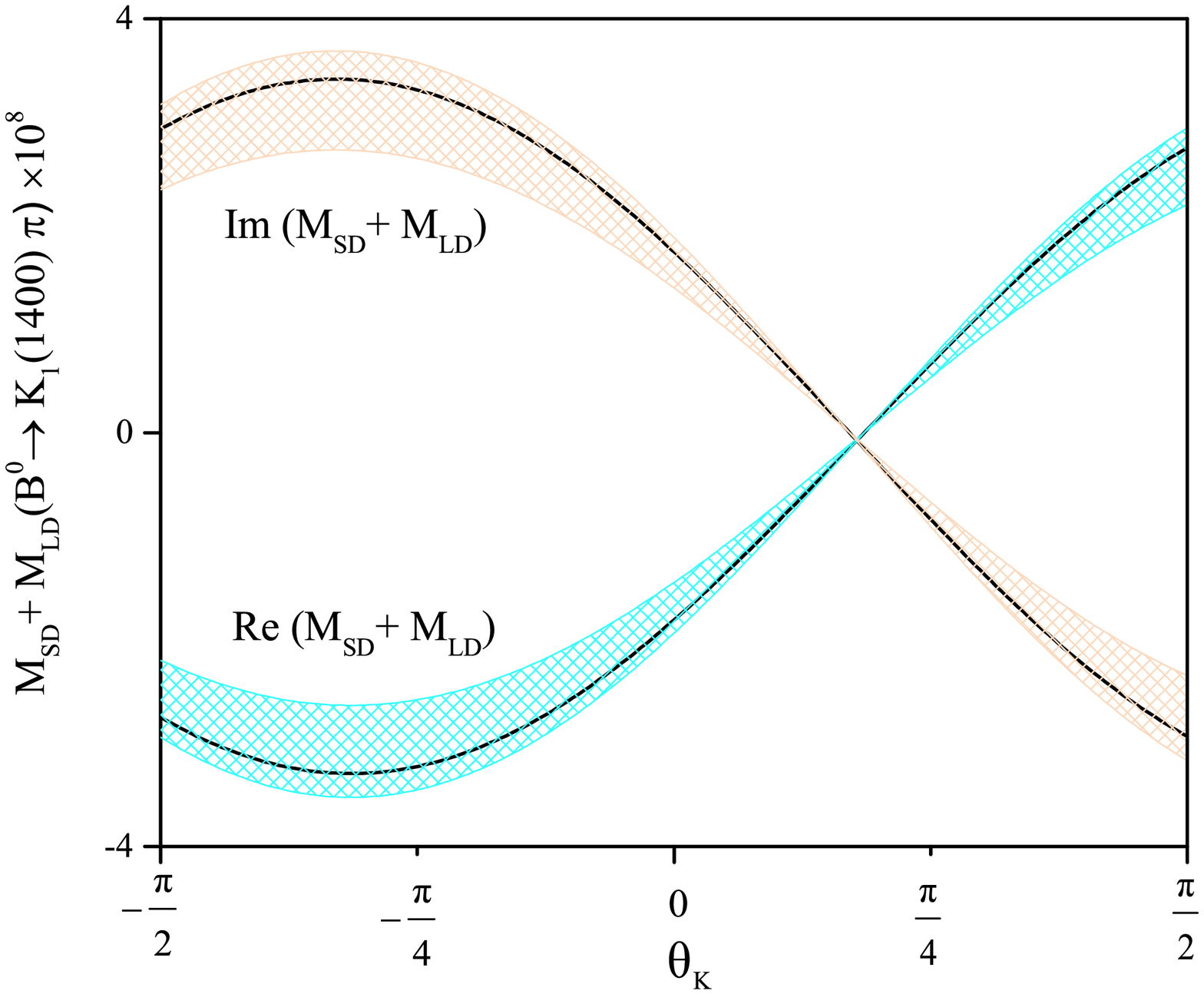}
\caption{ The same as Fig. \ref{sdamp} but for
$\mathcal{M}_{K_{1}\pi}=\mathcal{M}_{SD}+\mathcal{M}_{LD}$.}
\label{sdandtotalamp}
\end{figure}

In general,  ${\mathcal{M}}_{SD}$ is also a complex quantity like
${\mathcal{M}}_{LD}$. The input parameters in Eq. (\ref{eqsdam}) are
as $G_F=1.166\times 10^{-5}\, (\rm{GeV}^{-2})$, $V_{ub}^\ast
V_{us}=2.66\times 10^{-3}(1+2.90\, i)$, $V_{tb}^\ast
V_{ts}=-4.21\times 10^{-2}$ \cite{pdg}, $a_2=1.029 $, $a_4=0.005$,
$a_{10}=-0.001 $ \cite{Isola1}, $F_1^{B\to \pi}(m^2_{K_{1A}})=0.32
$, $F_1^{B\to \pi}(m^2_{K_{1B}})=0.33 $ \cite{Ali1998}.   First, we
plot the real and imaginary parts of $\mathcal{M}_{SD}$ for  $B^0
\to K_1^+ (1270,1400) \pi^-$ decays as a function of the mixing
angle $\theta_{K}$ in Fig. \ref{sdamp}.
Then,  to illustrate the impact of the LD in the amplitudes of $B^0
\to K_1^+ (1270,1400) \pi^-$ decays, we plot  the same as Fig.
\ref{sdamp} but for
$\mathcal{M}_{K_{1}\pi}=\mathcal{M}_{SD}+\mathcal{M}_{LD}$ in Fig.
\ref{sdandtotalamp}.

\begin{figure}
\includegraphics[width=7cm,height=6cm]{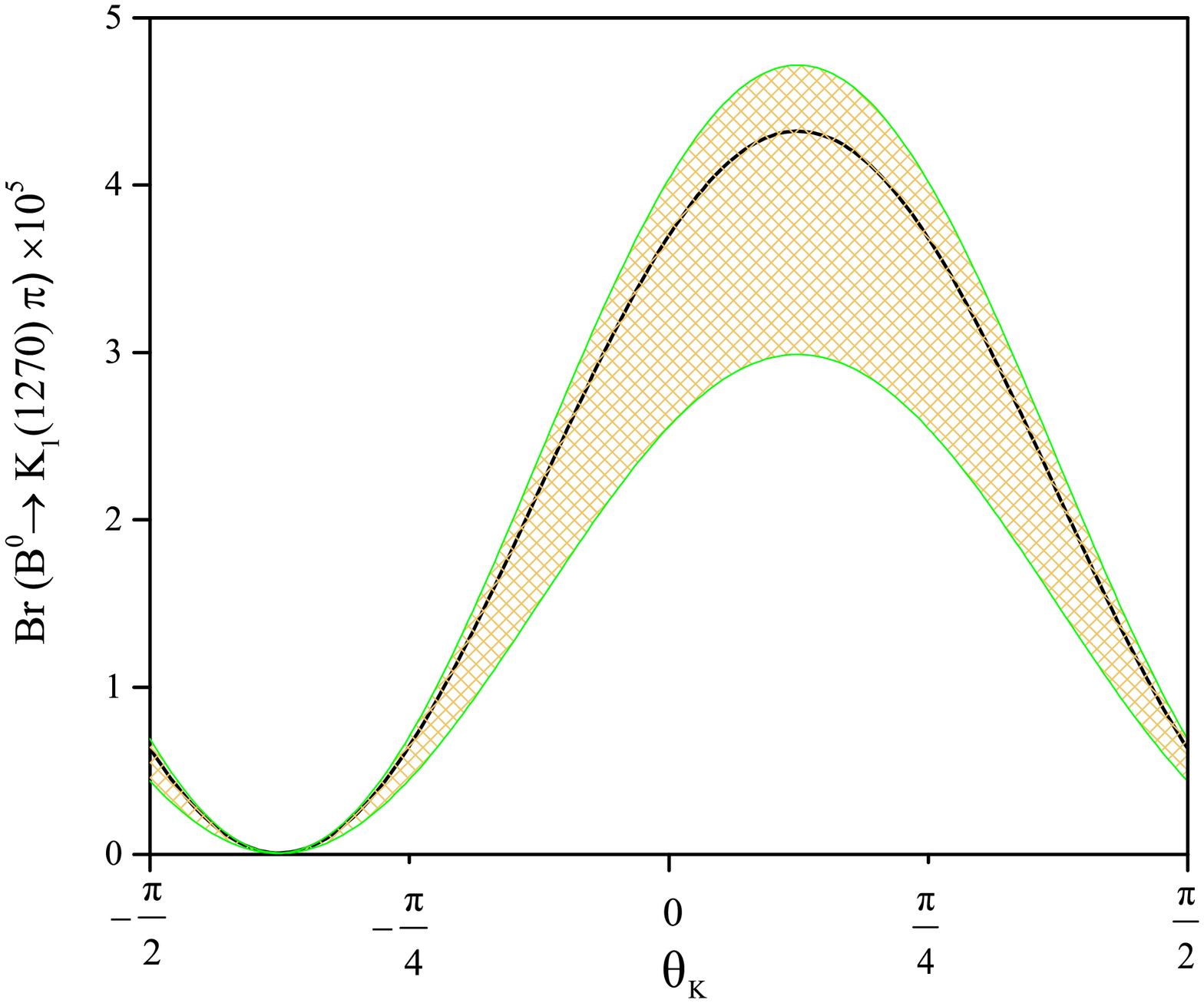}
\includegraphics[width=7cm,height=6cm]{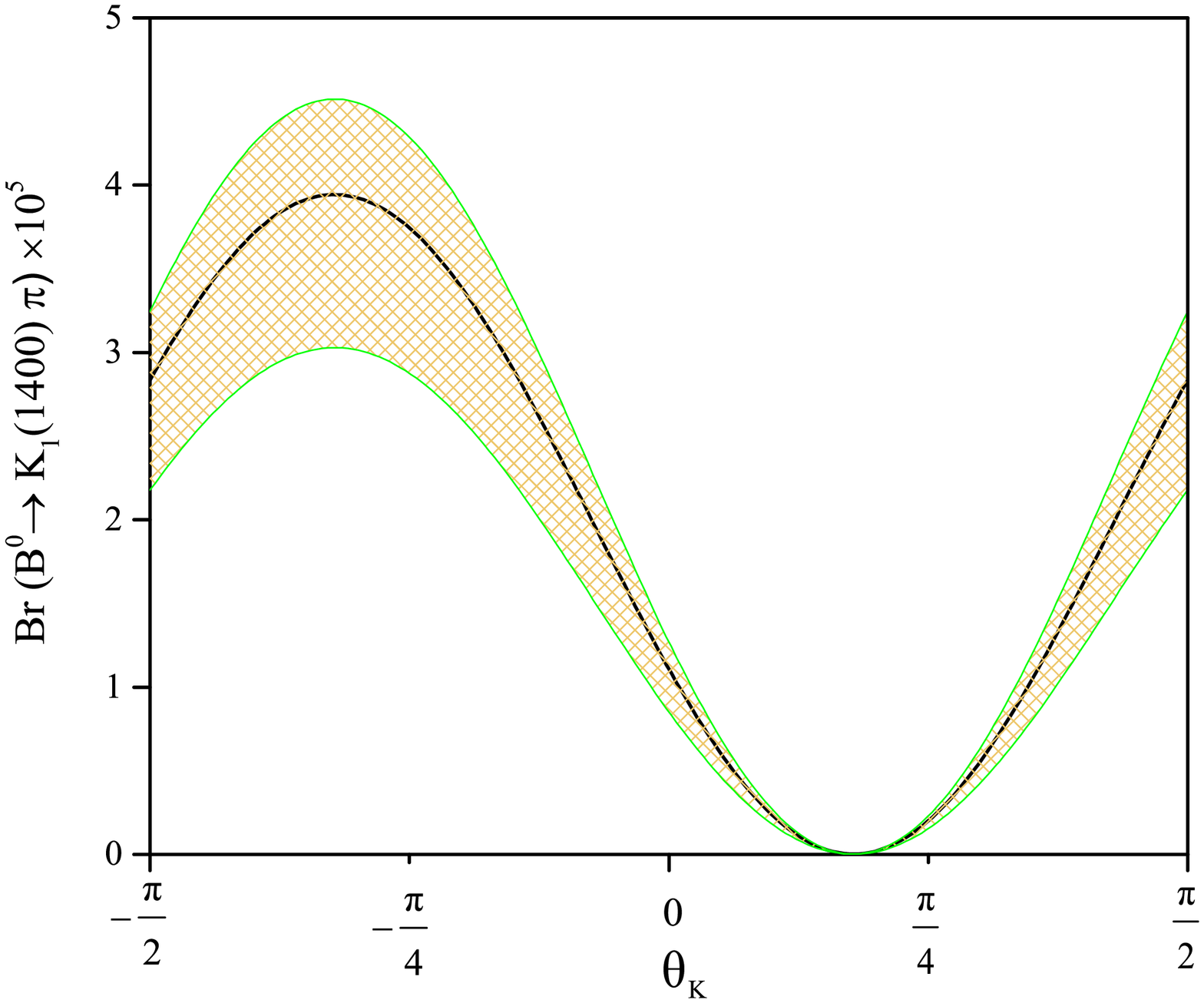}
\caption{Branching ratios for the  non-leptonic decays $B^0\to
{K}_{1}^{+}(1270) \pi^-$ and $B^0\to {K}_{1}^{+}(1400) \pi^-$ with
respect to $\theta_{K}$ and their uncertainty regions.} \label{br}
\end{figure}

Having $\mathcal{M}_{K_{1}\pi}$, the branching ratio of the
non-leptonic decay $B^0\to {K}_{1}^{+} \pi^-$ is given by
\begin{eqnarray}\label{eqbrtotal}
\mathcal{BR}(B^0\to {K}_{1}^{+} \pi^-) =\frac{\tau_B}{16 \pi
m^3_B}|\mathcal{M}_{K_{1}\pi}|^2 \sqrt{\lambda(m_B^2, m_{K_{1}}^2,
m_{\pi}^2)},
\end{eqnarray}
where $\tau_B$ is the life time of $B^0$ meson and $\lambda (m_B^2,
m_{K_{1}}^2,
m_{\pi}^2)=m_B^4+m_{K_{1}}^4+m_{\pi}^4-2m_B^2m_{K_{1}}^2-2m_B^2m_{\pi}^2-2m_{K_{1}}^2m_{\pi}^2$.
The $\theta_{K}$ dependence of the branching ratios for the
non-leptonic decays $B^0\to {K}_{1}^{+}(1270) \pi^-$ and $B^0\to
{K}_{1}^{+}(1400) \pi^-$ with their uncertainty regions is shown in
Fig. \ref{br}.

As mentioned before, the recent experimental values for the
$B^{0}\to K_{1}^{+}(1270, 1400) \pi^{-}$ branching ratios are
reported at $\theta_{K}=(72\pm3)^{\circ}$ by BABAR
\cite{Aubert2010}. Considering the amplitudes of $B^0 \to K_1^+
(1270,1400) \pi^-$ decays in different ways such as only
$(\mathcal{M}_{SD})$, $(\mathcal{M}_{SD}+\mathcal{M}_{LD}^{a+b})$,
$(\mathcal{M}_{SD}+\mathcal{M}_{LD}^{c})$,
$(\mathcal{M}_{SD}+\mathcal{M}_{LD}^{d+e})$, and  the total
amplitude $(\mathcal{M}_{SD}+\mathcal{M}_{LD})$, we present our
results for the branching ratios at $\theta_{K}=(72\pm3)^{\circ}$ in
Table \ref{Tabr}. According to the obtained values,
($\mathcal{M}_{SD}+\mathcal{M}_{LD}^{a+b}$) has the most
contribution to our results. The experimental values for these
considered decays are also reported in Table \ref{Tabr}. As can be
seen, our results are in a good agreement with the experimental
values.
\begin{table}[th]
\caption{ Branching ratio results for $B^0\to {K}_{1}^{+}(1270) \pi^-$ and $B^0\to {K}_{1}^{+}(1400) \pi^-$ decays at $\theta_{K}=(72\pm 3)^{\circ}$,
in units of $10^{-5}$.} \label{Tabr}
\begin{ruledtabular}
\begin{tabular}{ccccccc}
Decay & Only $\mathcal{M}_{SD}$&$\mathcal{M}_{SD}+\mathcal{M}_{LD}^{a+b}$&$\mathcal{M}_{SD}+\mathcal{M}_{LD}^{c}$&$\mathcal{M}_{SD}+\mathcal{M}_{LD}^{d+e}$&$\mathcal{M}_{SD}+\mathcal{M}_{LD}$&Exp \cite{Aubert2010}\\
\hline
 $B^0 \to K_{1}^{+}(1270) \pi^{-}$ &$0.26 \pm 0.04$&$1.19\pm0.11$  &$0.53\pm0.06$  &$0.11\pm0.02$  &$1.83 \pm 0.20 $&$1.7 \pm 0.4$\\
 $B^0 \to K_{1}^{+}(1400) \pi^{-}$&$0.20 \pm 0.03$&$1.03\pm0.10$  & $0.52 \pm0.05$ & $0.08\pm0.02$ &  $1.63 \pm 0.17$&$1.6 \pm 0.3$\\
\hline
Sum&$0.46 \pm 0.07$&$2.22\pm0.21$ &$1.05\pm0.11$
&$0.19\pm0.04$ & $3.46\pm 0.37$&$3.2 \pm 0.3$
\end{tabular}
\end{ruledtabular}
\end{table}

For a better analysis, other theoretical predictions  for the
branching ratios of $B^0\to K^+_1(1270,1400) \pi^-$ decays are also
presented in Table \ref{TREF}. It is noticed that the results of
Refs. \cite{CaCaVe,LaNaPh}  are obtained for mixing angle
$32^{\circ}$, while those in Ref. \cite{ChYaYa} are obtained for
mixing angle $-37^{\circ}$. Also, the mixing angle $\theta_{K}$ is
considered in two values $-33^{\circ}$ and $-58^{\circ}$ in Ref.
\cite{ZhHoYa}.
\begin{table}
\caption{Branching ratio values for $B^0\to K^+_1(1270,1400) \pi^-$
decays via other methods in different mixing angles in units of
$10^{-5}$. }\label{TREF}
\begin{ruledtabular}
\begin{tabular}{ccccccc}
Decay & $(\theta_K=32^{\circ})$\cite{CaCaVe}  & $(\theta_K=32^{\circ})$\cite{LaNaPh} &$(\theta_K=-37^{\circ})$\cite{ChYaYa} & $(\theta_K=-33^{\circ})$\cite{ZhHoYa}& $(\theta_K=-58^{\circ})$\cite{ZhHoYa}\\
\hline $ B^0\to
K^{+}_1(1270)\pi^-$&$0.43$&$0.76$&$0.30$&$0.46$&$0.32$\\
$ B^0\to  K^{+}_1(1400)\pi^-$ &$0.23$&$0.40$&$0.54$&$0.30$&$0.45$\\
\end{tabular}
\end{ruledtabular}
\end{table}
For a comparison, we show our results for the $B^{0}\to
K_{1}^{+}(1270, 1400) \pi^{-}$ branching ratios in the different
values of the mixing angle $\theta_{K}$ in Table \ref{TREF2}.
\begin{table}
\caption{Our results for the branching ratios of $B^0\to
K^+_1(1270,1400) \pi^-$ decays in units of $10^{-5}$. }\label{TREF2}
\begin{ruledtabular}
\begin{tabular}{ccccccc}
Decay &   $\theta_K=32^{\circ}$ &  $\theta_K=-37^{\circ}$& $\theta_K=-33^{\circ}$& $\theta_K=-58^{\circ}$ \\
\hline $ B^0\to
K^{+}_1(1270)\pi^-$&$4.19\pm0.85$&$1.54\pm 0.56$&$1.41\pm 0.34$&$0.12\pm 0.05$\\
$ B^0\to  K^{+}_1(1400)\pi^-$&$0.20\pm 0.02$&$3.42\pm 1.03$&$3.24\pm 0.43$&$3.94\pm 0.72$\\
\end{tabular}
\end{ruledtabular}
\end{table}
As can be seen in Tables \ref{TREF} and \ref{TREF2}, the values
predicted by us in the different angles are in most cases greater
than that those predicted by the other methods.

In summary, the strong coupling constants of $D D A$, $D^{*}D^{*}A$
and $D^{*} D A$ vertices were considered in the framework of the
LCSR, where $A$ is an axial vector meson such as $a_1, b_1, K_{1A},
K_{1B}, K_1(1270)$ and $K_1(1400)$. The branching ratio of the
non-leptonic decay $B^0\to {K}_{1}^{+}\pi^-$ was analyzed by using
the strong coupling constants of  $D_s D K_1$, $D_s^* D K_1$ and
$D_s^* D^* K_1$ vertices for $K_1(1270)$ and $K_1(1400)$ mesons. We
estimated the branching ratio values of these decays in different
values of the mixing angle $\theta_K$. Our results for the branching
ratios of $B^0\to {K}_{1}^{+}(1270,1400) \pi^-$ decays were in a
good agreement with the experimental values in $\theta_{K}=(72\pm
3)^{\circ}$.

\clearpage
\appendix
\begin{center}
{\Large \textbf{Appendix: Twist Function Definitions}}
\end{center}

In this appendix, we present the definitions for the two-parton
distribution amplitudes as well as the twist functions.

The two-parton chiral--even distribution amplitudes are given by
\cite{Kwei}:
\begin{eqnarray*}
\langle 0|\bar{q}(x) \gamma_\mu \gamma_5
q'(0)|A(p, \varepsilon)\rangle &=& i f_{A} m_{A}\int_0^1 du \,  e^{-i u
p. x} \Bigg\{ p_\mu \frac{\varepsilon. x}{p. x}
\Phi_\parallel(u) +\left( \varepsilon_{\mu} -p_\mu
\frac{\varepsilon. x}{p.x}\right) g_\perp^{(a)}(u)
+{\cal O}(x^2) \Bigg\},\nonumber\\
\langle 0|\bar{q}(x) \gamma_\mu q'(0)|A (p, \varepsilon)\rangle
& = & - i f_{A}\, m_{A}~
\epsilon_{\mu\nu\rho\sigma} \varepsilon^{\nu} p^{\rho} x^\sigma
\int_0^1 du \, e^{-i u \, p. x}\Bigg\{
\frac{g_\perp^{(v)}(u)}{4}+{\cal O}(x^2)\Bigg\},
\end{eqnarray*}
also, the two-parton chiral--odd distribution amplitudes are defined
by:
\begin{eqnarray*}
\langle 0|\bar{q}(x) \sigma_{\mu\nu}\gamma_5 q'(0) |A (p,
\varepsilon)\rangle & =&  f_{A}^{\perp} \int_0^1 du \, e^{-i u p'.
x} \Bigg\{(\varepsilon_{\mu} p_{\nu} - \varepsilon_{\nu} p_{\mu})
\Phi_\perp(u) + \frac{{m^2_{A}}\,\varepsilon. x}{(p. x)^2}(p_\mu
x_\nu - p_\nu x_\mu) \bar{h}_\parallel^{(t)}
+{\cal O}(x^2)\Bigg\}, \nonumber \\
\langle 0|\bar{q}(x) \gamma_5 q'(0) |A(p, \varepsilon)\rangle
&=& f_{A}^\perp m_{A}^2 (\varepsilon. x)\int_0^1 du
\, e^{-i u p. x}\Bigg\{\frac{h_\parallel^{(p)}(u)}{2}+ {\cal
O}(x^2)\Bigg\}.
\end{eqnarray*}

We take into account the approximate forms of the twist-2 functions,
for $A=a_1$ and $K_{1A}$ states, to be \cite{Kwei2}
\begin{eqnarray*}
\Phi_\parallel(u) & = & 6 u \bar u \left[ 1 + 3 a_1^\parallel\, \xi
+ a_2^\parallel\, \frac{3}{2} ( 5\xi^2  - 1 )
\right],\\
\Phi_\perp(u) & = & 6 u \bar u \left[ a_0^\perp + 3 a_1^\perp\, \xi
+ a_2^\perp\, \frac{3}{2} ( 5\xi^2  - 1 ) \right],
\end{eqnarray*}
and for $A=b_1$ and $K_{1B}$ to be
\begin{eqnarray*}
\Phi_\parallel(u) & = & 6 u \bar u \left[ a_0^\parallel + 3
a_1^\parallel\, \xi +
a_2^\parallel\, \frac{3}{2} ( 5\xi^2  - 1 ) \right],\\
\Phi_\perp(u) & = & 6 u \bar u \left[ 1 + 3 a_1^\perp\, \xi +
a_2^\perp\, \frac{3}{2} ( 5\xi^2  - 1 ) \right],
\end{eqnarray*}
where $\xi=2u-1$ and $\bar u= 1-u$. Also $a_i^\parallel$ and
$a_i^\perp$ ($i=0,1,2)$ are defined as the Gegenbauer moments of
$\Phi_\parallel$ and $\Phi_\perp$, respectively. The values of the
Gegenbauer moments for each axial vector meson are given in Ref.
\cite{Kwei}.

For the relevant two-parton twist-3 chiral-even distribution
amplitudes of $A=a_1$ and $K_{1A}$, we take the approximate
expressions up to conformal spin $9/2$ and ${\cal O}(m_s)$ as
\cite{Kwei2}:
\begin{eqnarray*}
g_\perp^{(a)}(u) & = &  \frac{3}{4}(1+\xi^2) + \frac{3}{2}\,
a_1^\parallel\, \xi^3 + \left(\frac{3}{7} \, a_2^\parallel + 5
\zeta_{3, A}^V \right) \left(3\xi^2-1\right)
\nonumber\\
& & {}+ \left( \frac{9}{112}\, a_2^\parallel + \frac{105}{16}\,
\zeta_{3, A}^A - \frac{15}{64}\, \zeta_{3, A}^V \omega_{A}^V
\right) \left( 35\xi^4 - 30 \xi^2 + 3\right) \nonumber\\
& & + 5\Bigg[ \frac{21}{4}\zeta_{3, A}^V \sigma_{A}^V + \zeta_{3,
A}^A \bigg(\lambda_{A}^A -\frac{3}{16} \sigma_{A}^A\Bigg)
\Bigg]\xi(5\xi^2-3)
\nonumber\\
& & {}-\frac{9}{2} \bar{a}_1^\perp
\,\widetilde{\delta}_+\,\left(\frac{3}{2}+\frac{3}{2}\xi^2+\ln u
+\ln\bar{u}\right) - \frac{9}{2}
\bar{a}_1^\perp\,\widetilde{\delta}_-\, ( 3\xi + \ln\bar{u} - \ln
u),
\end{eqnarray*}
\begin{eqnarray*}
g_\perp^{(v)}(u) & = & 6 u \bar u \Bigg\{ 1 + \Bigg(a_1^\parallel +
\frac{20}{3} \zeta_{3, A}^A
\lambda_{A}^A\Bigg) \xi\nonumber\\
&& + \Bigg[\frac{1}{4}a_2^\parallel + \frac{5}{3}\, \zeta^V_{3, A}
\left(1-\frac{3}{16}\, \omega^V_{A}\right)
+\frac{35}{4} \zeta^A_{3, A}\Bigg] (5\xi^2-1) \nonumber\\
&&+ \frac{35}{4}\Bigg(\zeta_{3, A}^V \sigma_{A}^V
-\frac{1}{28}\zeta_{3, A}^A
\sigma_{A}^A \Bigg) \xi(7\xi^2-3) \Bigg\}\nonumber\\
& & {} -18 \, \bar{a}_1^\perp\widetilde{\delta}_+ \,  (3u \bar{u} +
\bar{u} \ln \bar{u} + u \ln u ) - 18\,
\bar{a}_1^\perp\widetilde{\delta}_- \,  (u \bar{u}\xi + \bar{u} \ln \bar{u} -
u \ln u),
\end{eqnarray*}
\begin{eqnarray*}
h_\parallel^{(t)}(u) &= & 3a_0^\perp\xi^2+ \frac{3}{2}\,a_1^\perp
\,\xi (3 \xi^2-1) + \frac{3}{2} \Bigg[a_2^\perp \xi +
\zeta^\perp_{3, A}\Bigg(5
-\frac{\omega_{A}^{\perp}}{2}\Bigg)\Bigg]\, \xi \,(5\xi^2-3)
\nonumber\\
&& +\frac{35}{4}\zeta^\perp_{3, A} \sigma^\perp_{A}
(35\xi^4-30\xi^2+3) + 18 \bar{a}_2^\parallel
\Bigg[\delta_+ \xi -\frac{5}{8}\delta_- (3\xi^2-1)\Bigg]\nonumber\\
&& - \frac{3}{2}\, \Bigg( \delta_+ \, \xi [2 +  \ln (\bar{u}u)]
+\,\delta_- \, [ 1 + \xi \ln (\bar{u}/u) ]\Bigg) (1+ 6
\bar{a}_2^\parallel),
\end{eqnarray*}
\begin{eqnarray*}
h_\parallel^{(p)}(u) & = & 6u\bar u \Bigg\{ a_0^\perp +
\Bigg[a_1^\perp +5\zeta^\perp_{3, A}\Bigg(1-\frac{1}{40}(7\xi^2-3)
\omega_{A}^{\perp} \Bigg)\Bigg] \xi \nonumber\\
&& \ \ \ \ \ \ \ + \Bigg( \frac{1}{4}a_2^\perp + \frac{35}{6}
\zeta^\perp_{3, A} \sigma^\perp_{A} \Bigg) (5\xi^2-1)
-5\bar{a}_2^\parallel \Bigg[\delta_+ \xi + \frac{3}{2} \delta_-
(1-\bar{u} u) \Bigg]\Bigg\}
\nonumber\\
& & {}- 3[\, \delta_+\, (\bar{u} \ln \bar{u} - u \ln u) +
\,\delta_-\,  ( u \bar{u} + \bar{u} \ln \bar{u} + u \ln u)] (1+ 6
\bar{a}_2^\parallel),
\end{eqnarray*}
where
\begin{equation*}
\widetilde{\delta}_\pm  ={f_{A}^{\perp}\over f_{A}}{m_{q_2} \pm
m_{q_1} \over m_{A}},\qquad \zeta_{3,A}^{V(A)} =
\frac{f^{V(A)}_{3A}}{f_{A} m_{A}}.
\end{equation*}
In these phrases, the values of all parameters such as
$\omega_{A}^V$, $\sigma^\perp_{A}$ and etc.  are given in Ref.
\cite{Kwei} for each meson.

On the other hand, the same as the above quantities but for $A=b_1$
and $K_{1B}$ states are given as follows.
\begin{eqnarray*}
g_\perp^{(a)}(u) & = & \frac{3}{4} a_0^\parallel (1+\xi^2) +
\frac{3}{2}\, a_1^\parallel\, \xi^3 + 5\left[\frac{21}{4}
\,\zeta_{3, A}^V + \zeta_{3, A}^A
\Bigg(1-\frac{3}{16}\omega_{A}^A\Bigg)\right]
\xi\left(5\xi^2-3\right)
\nonumber\\
& & {}+ \frac{3}{16}\, a_2^\parallel \left(15\xi^4 -6 \xi^2
-1\right) + 5\, \zeta^V_{3, A}\lambda^V_{A}\left(3\xi^2 -1\right)
\nonumber\\
& & {}+ \frac{105}{16}\left(\zeta^A_{3, A}\sigma^A_{A}
-\frac{1}{28} \zeta^V_{A}\sigma^V_{A}\right)
\left(35\xi^4 -30 \xi^2 +3\right)\nonumber\\
& & {}-15\bar{a}_2^\perp \bigg[ \widetilde{\delta}_+ \xi^3 +
\frac{1}{2}\widetilde{\delta}_-(3\xi^2-1) \bigg] \nonumber\\
& & {} -\frac{3}{2}\,\bigg[\widetilde{\delta}_+\, ( 2 \xi +
\ln\bar{u} -\ln u) +\, \widetilde{\delta}_-\,(2+\ln u +
\ln\bar{u})\bigg](1+6\bar{a}_2^\perp),
\end{eqnarray*}
\begin{eqnarray*}
g_\perp^{(v)}(u) & = & 6 u \bar u \Bigg\{ a_0^\parallel +
a_1^\parallel \xi + \Bigg[\frac{1}{4}a_2^\parallel +\frac{5}{3}
\zeta^V_{3, A} \Bigg(\lambda^V_{A} -\frac{3}{16} \sigma^V_{A}\Bigg)
+\frac{35}{4} \zeta^A_{3, A}\sigma^A_{A}\Bigg](5\xi^2-1) \nonumber\\
& & {}  + \frac{20}{3}\,  \xi \left[\zeta^A_{3, A} + \frac{21}{16}
\Bigg(\zeta^V_{3, A}- \frac{1}{28}\, \zeta^A_{3, A}\omega^A_{A}
\Bigg)
(7\xi^2-3)\right]\nonumber\\
& & {} -5\, \bar{a}_2^\perp [2\widetilde\delta_+ \xi +
\widetilde\delta_- (1+\xi^2)]
\Bigg\}\nonumber\\
& & {} - 6 \bigg[\, \widetilde{\delta}_+ \, (\bar{u} \ln\bar{u}
-u\ln u ) +\, \widetilde{\delta}_- \, (2u \bar{u} + \bar{u} \ln
\bar{u} + u \ln u)\bigg] (1+6\bar{a}_2^\perp) ,
\end{eqnarray*}
\begin{eqnarray*}
h_\parallel^{(t)}(u) &= & 3\xi^2+ \frac{3}{2}\,a_1^\perp \,\xi
(3\xi^2-1) + \Bigg[\frac{3}{2} a_2^\perp\, \xi +
\frac{15}{2}\zeta^\perp_{3, A} \Bigg(\lambda^\perp_{A} -
\frac{1}{10} \sigma^\perp_{A}\Bigg) \Bigg]
\, \xi(5\xi^2-3) \nonumber\\
&& {} +\frac{35}{4}\zeta^\perp_{3, A}(35\xi^4-30\xi^2+3)\nonumber\\
& & {} +\frac{9}{2} \bar{a}_1^\parallel\, \xi \Bigg[\delta_+\, (\ln
u - \ln \bar{u} -3\xi) - \delta_-\,  \Bigg( \ln u + \ln \bar{u}
+\frac{8}{3}\Bigg)\Bigg]\,,
\end{eqnarray*}
\begin{eqnarray*}
h_\parallel^{(p)}(u) & = & 6u\bar u \Bigg\{ 1 + a_1^\perp \xi +
\left(\frac{1}{4}a_2^\perp + \frac{35}{6}\,\zeta^\perp_{3, A}
\right)(5\xi^2-1)
\nonumber\\
& & {} +5\zeta^\perp_{3, A}
\Bigg[\lambda^\perp_{A}-\frac{1}{40}(7\xi^3-3)\sigma^\perp_{A}
\Bigg] \, \xi
\Bigg\}  \nonumber\\
& & {} -9\bar{a}_1^\parallel\, \delta_+\, (3 u \bar{u} + \bar{u} \ln
\bar{u} + u \ln u) -9\bar{a}_1^\parallel\,\delta_-\,  \Bigg(
\frac{2}{3}\xi u\bar{u} + \bar{u} \ln \bar{u} - u \ln u \Bigg)\,.
\end{eqnarray*}

\end{document}